\documentclass[fleqn,usenatbib]{mnras}

\usepackage{newtxtext,newtxmath}

\usepackage[T1]{fontenc}
\usepackage[dvipsnames]{xcolor}

\DeclareRobustCommand{\VAN}[3]{#2}
\let\VANthebibliography\thebibliography
\def\thebibliography{\DeclareRobustCommand{\VAN}[3]{##3}\VANthebibliography}

\usepackage{graphicx}	
\usepackage{amsmath}	

\usepackage{subfiles}
\usepackage{xspace} 

\newcommand{\oii}{\ifmmode [\mathrm{O}\,\textsc{ii}] \else [O~{\sc ii}]\fi\xspace}

\newcommand{\oiii}{\ifmmode [\mathrm{O}\,\textsc{iii}] \else [O~{\sc iii}]\fi\xspace}

\newcommand{\caii}{\ifmmode \mathrm{Ca}\,\textsc{ii}\,\mathrm{(H\&K)} \else Ca~{\sc ii}~(H\&K)\fi\xspace}

\newcommand{\mgi}{\ifmmode \mathrm{Mg}\,\textsc{i b} \else Mg~{\sc i}\fi\xspace}

\newcommand{\hii}{\ifmmode \mathrm{H}\,\textsc{ii} \else H~{\sc ii}\fi\xspace}

\newcommand{\neiii}{\ifmmode [\mathrm{Ne}\,\textsc{iii}] \else [Ne~{\sc iii}]\fi\xspace}

\newcommand{\nev}{\ifmmode [\mathrm{Ne}\,\textsc{v}] \else [Ne~{\sc v}]\fi\xspace}

\title[Host galaxies of quasars]{Stellar populations of quasar host galaxies with MFICA decomposition}

\author[S. D. Krishna et al.]{
Sahyadri D. Krishna,$^{1}$\thanks{E-mail: sk322@st-andrews.ac.uk (SDK)}
Vivienne Wild,$^{1}$
Paul C. Hewett$^{2}$,
Carolin Villforth$^{3}$
\\
$^{1}$SUPA, School of Physics and Astronomy, University of St Andrews, North Haugh, St Andrews KY16 9SS, UK\\
$^{2}$Institute of Astronomy, University of Cambridge, Madingley Road, Cambridge CB3 0HA, UK\\
$^{3}$Department of Physics, University of Bath, Claverton Down, Bath BA2 7AY, UK
}

\date{Accepted XXX. Received YYY; in original form ZZZ}

\pubyear{2025}

\begin{document}
\label{firstpage}
\pagerange{\pageref{firstpage}--\pageref{lastpage}}
\maketitle

\begin{abstract}
Galaxy evolution theories require co-evolution between accreting supermassive black holes (SMBH) and galaxies to explain many properties of the local galaxy population, yet observational evidence for the mechanisms driving this co-evolution is lacking. The recent star-formation histories of the host galaxies of accreting SMBHs (Active Galactic Nuclei, AGNs) can help constrain the processes that feed SMBHs and halt star formation in galaxies, but are difficult to obtain for the most luminous AGNs (quasars). We introduce Mean-Field Independent Component Analysis (MFICA) to decompose quasar spectra and obtain recent star formation histories of their host galaxies. Applying MFICA to quasar spectra from the Sloan Digital Sky Survey (SDSS) DR7 Quasar Catalogue in the redshift range $0.16 \leq z \leq 0.76$, we find that 53 per cent of quasar host galaxies are star-forming, 17 per cent lie in the green-valley, while only 5 per cent are quiescent. This contrasts with 14, 11, and 74 per cent of a mass-matched control sample that are star-forming, green-valley, and quiescent, respectively. We find that $\sim25$ per cent of quasars are hosted by post-starburst galaxies, an excess of $28\pm1$ compared to our control sample. While the heterogeneity of recent star formation histories implies multiple SMBH feeding mechanisms, the excess of post-starburst host galaxies demonstrates the link between accreting SMBHs and a recent starburst followed by rapid quenching. Given that massive post-starburst galaxies are predominantly caused by gas-rich major mergers, our results indicate that $30-50$ per cent of quasars originate from merger-induced starbursts.
\end{abstract}




\section{Introduction}

A central challenge in extragalactic astronomy is characterising the mechanisms responsible for altering the rate of star formation in galaxies \citep{gabor2010} and reproducing the bi-modality of galaxy colours in the local universe \citep{baldry2004}. Supermassive black holes (SMBHs) hosted by most massive galaxies are implicated to play a leading role. Scaling relations between SMBH and galaxy properties, such as the $M_{\rm BH}-\sigma_{*}$ \citep{ferrarese2000, gebhardt2000} or $M_{\rm BH}-M_{\rm bulge}$ \citep{magorrian1998,haring2004} relations, suggest that galaxies and black holes evolve in step \citep{silk1998, kormendy2013} or that a common phenomenon regulates their growth and evolution \citep{jahnke2011}. 

SMBHs can accrete gas from their surroundings, forming accretion disks that radiate across the electromagnetic spectrum in episodes known as Active Galactic Nuclei (AGNs). Numerical simulations require some fraction of energy from an AGN to couple to the gas in a galaxy -  a phenomenon known as AGN feedback \citep{silk1998,fabian2012} - to prevent the overproduction of stars in massive galaxies \citep{benson2003}, reproduce observed bimodality \citep{croton2006}, and modulate the growth of black holes \citep{dimatteo2005}. 

However, conclusive evidence for AGN feedback is missing, and it is unclear which galaxy scale processes (e.g., mergers, bars, disk instabilities) are most important in triggering AGNs and growing black holes \citep[see][for reviews on AGN feedback and feeding]{fabian2012,alexander2012}. The crucial evidence may lie in studying the star-formation histories (SFHs) of AGN host galaxies, which connect changes in galaxy-scale gas supplies to black hole accretion. 

Studies on the ongoing star-formation properties of AGN hosts have returned mixed results. Some studies find that AGNs are hosted by star-forming galaxies on the main sequence of star-formation \citep{Silverman2009, Mullaney2012,Trump2013, Rosario2013, Yue2018, Zhuang2022} with blue colours \citep{Li2021} and young stellar populations \citep{Kauffmann2003agn, holt2007, bessiere2017, burtscher2021,  Ren2024}. Active star formation in AGN hosts might suggest that AGN feedback is absent/inefficient, or that AGN activity promotes star-formation \citep{zinn2013}. Alternatively, active star-formation may 
manifest from the requirement of cold gas for both AGN activity and star formation or due to star-formation fueling SMBH accretion \citep{ciotti2007, kauffmann2009, Ni2023}. Star-forming AGN hosts may also be the sites of AGN feedback in the future \citep{bar2017, harrison2024}. In fact, \citet{Ward2022} show that AGNs reside in star-forming galaxies even in simulations as the impact of AGN feedback is delayed.

Contrary to the finding of star-forming hosts, many other studies suggest that AGN hosts lie in the “green valley” with star-formation rates (SFRs) below the star-formation main sequence \citep{salim2007, Silverman2008, Silverman2009, shimizu2015, leslie2015, ellison2016}, green colours \citep{martin2007, Schawinski2009} and older stellar populations \citep{matsuoka2015}. The presence of AGN host galaxies in the green valley might implicate the AGN in causing the transition of galaxies from star formation to quiescence \citep{Silverman2008, leslie2015}. However, some studies find a delay between the end of star formation and AGN activity \citep{davies2007, Schawinski2007, Wild2010}, implying that AGN activity is merely coincident and not responsible for the quiescence of a galaxy \citep{Hopkins2012, Lanz2022}. Finally, several studies, including some of the earliest studies of AGN host galaxy properties, find AGN hosts to be quiescent \citep{mclure1999, nolan2001, birchall2023, Ni2023}.

A range of factors complicates the consolidation and interpretation of these results. AGNs vary over timescales shorter than changes in star formation in galaxies \citep{novak2011}. This variability and the short duty cycles of AGNs, which result in AGNs turning on and off over many cycles \citep[e.g.][]{shankar2009}, can make it challenging to draw physical connections between AGN activity and host galaxy properties, even when these connections are present \citep{aird2019}. AGN host galaxies also show different properties when selected in different wavelengths  \citep{Hickox2009, ellison2016, ji2022}. Finally, care must be taken to account for the influence of latent physical properties, like stellar mass, which can explain differences between the properties of AGN host galaxies and their inactive counterparts \citep{Silverman2009}.

Thus, a measured, statistical treatment of large samples of galaxy and AGN properties is needed to investigate AGN-galaxy co-evolution. Rest-frame optical spectra of galaxies are a rich source of information, with absorption and emission lines encoding their current stellar populations and past star-formation histories \citep[see][for a review on SFH modelling]{conroy2013}. Where feasible, stellar population modelling of AGN host galaxy spectra has revealed significant intermediate-age (lifetimes of $<1-2$\,Gyrs) stellar populations dominated by A and F-type stars \citep{Kauffmann2003agn, vdb2006, canalizo2013, Cales2013, cales2015, matsuoka2015, riffel2023}, whose presence (characterised by strong Balmer absorption) along with an absence of short-lived O and B stars is a characteristic of post-starburst, or E+A, galaxies \citep[see][for a review on post-starburst galaxies]{french2021}. Though rare in the local universe \citep{pawlik2018}, studies have shown that a large fraction of the progenitors of quiescent galaxies may have been rapidly quenched post-starburst galaxies \citep{yesuf2014, Wild2020, Taylor2023}. Numerical simulations widely predict the co-existence of luminous AGN activity - particularly a quasar phase - alongside post-starburst-like stellar populations \citep{hopkins2008, snyder2011, lotz2021}. Results on studying the coincidence of AGN activity and the post-starburst phase in observations, however, have been mixed \citep[][]{goto2006, yan2006, brown2009, yesuf2014, Lanz2022, almaini2025}. 

Studies of AGN host galaxy spectra have focussed primarily on low-luminosity Type-2 AGNs \citep[e.g.][]{Kauffmann2003agn} where a dusty torus is thought to block light from the accretion disk of the AGN, allowing modelling of the host galaxy continuum. These studies have then invoked the unified model of AGNs to argue that their results are generalisable to all AGNs. Such generalisations were adopted owing to the challenge of modelling the host galaxies of more luminous Type-1 AGNs and quasars, where an unobscured accretion disk outshines the host galaxy. The host galaxy properties of Type-1 quasars, in particular their stellar populations, are thus not well constrained, and it is unclear if their properties are the same as other less-luminous Type-2 AGNs, or perhaps different \citep{kim2006, Trump2013, chen2015, zakamska2016, zou2019, suh2019, Zhuang2022}. Special ``decomposition'' techniques are required to disentangle the AGN and host galaxy contributions embedded in the spectra of Type-1 AGNs and quasars, and subsequently enable modelling of their host stellar populations. 

\subsection{Decomposing AGN spectra}

Several decomposition techniques have been applied to observations of a single unresolved spectrum, such as the spectra from the Sloan Digital Sky Survey \citep[SDSS;][]{york2000}. These decomposition techniques rely on simultaneously fitting templates - either theoretical models or empirical templates - that individually describe AGN and galaxy spectra \citep[e.g.][]{buchner2024}. Stellar population synthesis (SPS) models are often used for the host galaxy \citep[e.g.][]{matsuoka2015, cales2015, wu2018} as they enable extraction of valuable physical properties, such as galaxy stellar mass, SFR, or stellar population age, directly from the decomposition. The contribution of the accreting black hole to the AGN spectrum can be modelled using parametric components (e.g., a power-law continuum) and/or empirical templates to reconstruct individual emission line strengths and line complexes \citep{temple2021}. Studies employing decomposition with theoretical models or empirical templates, however, are often limited in applicability to spectra with high signal-to-noise ratios \citep{matsuoka2015}, or to spectra pre-selected with significant galaxy starlight contribution \citep{cales2015, wu2018}.

Instead of using theoretical or empirical templates, many studies have used Principal Component Analysis (PCA) for decomposition. PCA has been used to derive orthogonal components to reconstruct quasar \citep{francis1992, yip2004qso} and galaxy \citep{connolly1995, ronen1999, yip2004gal, Wild2007} spectra, which are then combined in linear decomposition to decompose quasar spectra \citep{vdb2006, Shen2008, Shen2015, Jalan2023, Ren2024}. PCA decomposition has a major limitation: galaxy and AGN spectra are not naturally decomposed into statistically orthogonal components, and thus PCA produces components with features of different physical origins mixed together. Irrespective of the origin of the components, it is difficult to fully account for the contamination by residual levels of host galaxy light.

In this study, we introduce the use of Mean-Field Independent Component Analysis (or MFICA) as an alternative data-driven technique for decomposing quasar spectra from SDSS. The MFICA code we use in this paper was adapted from the codes of \citet{hojensorensen2002} and \citet{opper2005}, used to classify galaxies by \citet{Allen2013}, and reconstruct quasar spectra by \citet{Rankine2020}. In Section \ref{sample selection} we give a mathematical description of MFICA, describe the selection of training and testing samples for MFICA, spectrum pre-processing, and generation of templates with MFICA. In Section \ref{val_gal}, we explore properties of the MFICA galaxy templates. In Section \ref{validation}, we validate the performance of MFICA by decomposing mock spectra. In Section \ref{results}, we apply MFICA decomposition to a sample of quasar spectra from SDSS, and discuss the implications of the results in Section \ref{discussion}. Throughout this paper, we assume a $\rm \Lambda CDM$ cosmology, with $H_{0}=70\rm \,km\, s^{-1} Mpc^{-1}$, $\Omega_m=0.3$, and $\Omega_{\Lambda}=0.7$. If you are more interested in the science results of this paper, we recommend skipping the technical sections and going straight to Section \ref{results}.

\section{Mean Field Independent Component Analysis}\label{sample selection}

In Sections \ref{whatmfica} and \ref{objmfica}, we describe Mean-Field Independent Component Analysis (MFICA) and motivate its use. In Section \ref{mficatrain}, we describe the collation of training and testing samples of spectra for MFICA. Finally in Section \ref{makecomp} we describe our approach to generate templates capable of reconstructing the spectra of galaxies and high-luminosity quasars using MFICA.

\subsection{What is MFICA?}\label{whatmfica}

Independent Component Analysis, or ICA, is an extensively applied blind source separation (BSS) technique that also possesses dimensionality reduction capabilities analogous to PCA. Given some data X, BSS techniques try to estimate the original signals/sources, S, that were mixed together in unknown proportions to produce the observed data X. BSS can be mathematically formulated as finding the matrix of sources S, from the matrix of data X, such that:

\begin{equation}\label{mix}
X = AS
\end{equation}
where A is known as the mixing matrix.

Neither S nor A are known a priori. Thus, to solve equation (\ref{mix}), some very general assumption needs to be made about the sources S. In PCA, the sources S are required to be orthogonal. In ICA, the sources S are instead required to be statistically independent. ICA thus finds those sources that satisfy the inversion of equation (\ref{mix}) and are maximally statistically independent. Maximising independence, however, is not straightforward, and so most ICA codes maximise some proxy of independence \citep[see][for a review on ICA]{hyvarinen2001}. ICA has seen some use in source separation problems in astrophysics \citep{waldmann2013, Allen2013, querejeta2015}. It has also been used in the reconstruction of spectra, where the derived sources were found to be related to the stellar populations ages of galaxies \citep{kaban2005, lu2006}.

For observational data, such as SDSS spectra, equation (\ref{mix}) can be reformulated as:

\begin{align}\label{mix_noise}
X = AS + \Sigma
\end{align}
where $\Sigma$ is noise present in the original data X \citep{hojensorensen2002}. One approach to solving for the sources and other unknown quantities is to use Bayes Theorem, and invoke prior knowledge for the sources S \citep{hojensorensen2002}. A, S and $\Sigma$ can then be found by maximising likelihood or maximising a-posteriori probability \citep{hojensorensen2002}. The assumption of independence, however, is not sufficient to solve for the unknowns this way. The true posteriors of the sources are unknown a priori.

One way of circumventing this issue is to construct a ``good'' approximation to the unknown true posterior of the sources S. This can be achieved through a technique like mean-field variational theory, and is the premise for Mean-Field ICA \citep[MFICA; see][for a review]{hojensorensen2002}. 

\subsection{Objectives with MFICA}\label{objmfica}

We want to apply (or rather train) MFICA on samples of galaxy and quasar spectra to create a set of templates (known as components), capable of reconstructing galaxy and quasar spectra. The use of MFICA for the creation of spectral templates over other data-driven methods offers several benefits:

\begin{enumerate}
    \item MFICA does not require rescaling of data to unit variance and zero mean, i.e. no whitening \citep[see][for more on whitening and ICA]{stone2004}. Data whitening is counter-intuitive when the data is intrinsically composed of positive constituents, like stars in galaxy spectra. Whitening also distorts the natural shape of spectra, altering the form of features identified by template generation schemes.
    \item Like other Bayesian inference techniques, MFICA allows the incorporation of prior information on the original components S, such as a constraint of positivity. This positivity constraint was used by \citet{Allen2013} to generate galaxy templates and by \citet{Rankine2020} to generate quasar templates. The combination of positivity and un-whitened spectra results in positive components with easily inferred physical meaning.
    \item MFICA allows pre-specification/defining fixed components. This pre-specification and the ability to include strong priors on the sources simplifies the MFICA analysis. For example, MFICA can be applied in series to different galaxy subsamples, as we already know that galaxies are composed of different stellar populations. Differences in the relative fractions of stellar populations create distinct subsamples of galaxies. 
\end{enumerate}

\subsection{MFICA Training Samples}\label{mficatrain}

MFICA requires both samples of galaxy and quasar spectra for the creation of components capable of reconstructing those spectra. The construction of individual training samples of galaxies and quasars is described below.

\subsubsection{MFICA galaxy training sample}

Galaxies used for the creation of MFICA galaxy components were taken from the SDSS data release 7 \citep[DR7;][]{abazajian2009} main galaxy catalogue \citep{strauss2002}. We selected galaxies with secure spectroscopic redshifts in the range $0.16 < z < 0.38$ and mean per pixel spectral signal-to-noise ratio (S/N) $> 8$, calculated over a wavelength range of 3797-8400\,\AA. Next, using emission line properties in the MPA-JHU catalogue \citep{Brinchmann2004, tremonti2004}, we removed objects classified as AGNs or LINERs based on their emission line ratios (i.e. $i_{class}$ = 3, 4 and 5 in \citealt{Brinchmann2004}) and  dusty galaxies with Balmer decrements (the ratio of H$\alpha$ to H$\beta$ emission line fluxes) $> 4.8$, where each emission line is detected with S/N $> 5$. We also removed potential AGNs with ratio of velocity dispersions of Balmer to forbidden lines $> 1.6$, when all emission lines are detected with S/N ratio $> 5$. Finally, we removed a small number of objects with unphysical absolute $r$-band magnitudes and less than 3825 good pixels in their spectra. The remaining galaxy spectra make up the MFICA parent galaxy sample.

Like other artificial intelligence (AI) techniques, MFICA requires a training dataset that is unbiased but also large enough to quantify intrinsic features present in galaxy spectra. The construction of the ideal training sample thus requires quantifying the diversity of spectra in the parent sample. We do this by splitting the sample into different types using the principal component analysis (PCA) catalogue of \citet{Wild2007}, which provides information on the recent ($\leq 1$\,Gyr) star-formation histories of galaxies. Using the \citet{Wild2007} demarcation in PC1-PC2, we split the sample into quiescent, star-forming, starburst, post-starburst and green-valley galaxies. Selecting a similar number of galaxies in each class ensures that we encompass most of the variation observed in galaxy spectra, and no single galaxy class dominates the component generation. We limit the number of quiescent galaxies in the parent sample to 2000 to ensure balance in the numbers of the different galaxy types, obtaining a parent sample of 6875 galaxies. The final MFICA galaxy training sample, selected from the parent sample, consists of a random selection of 680 starburst galaxies, 1049 star-forming galaxies, 264 post-starburst galaxies, 554 green-valley galaxies and 823 red galaxies. This gives us a training sample of 3370 galaxies. 

\subsubsection{MFICA Quasar Training Sample}

Quasar spectra were taken from the \citet{schneider2010} SDSS DR7 Quasar Catalogue with redshifts calculated using the scheme of \citet{hewett2010}. We selected quasars in the redshift range $0.25 < z < 0.5$, with mean per pixel spectral S/N $> 8$, and $i$-band absolute magnitudes $M_{i} < -23$, a magnitude brighter than the limit for inclusion in the catalogue. The $M_{i}$ cut excludes most objects dominated by host-galaxy light, although we explicitly remove any remaining host-galaxy contamination in Section \ref{sec:QSOcomponents}. Apart from a weak systematic change in emission line equivalent-width with luminosity \citep[the Baldwin effect;][]{baldwin1997}, the spectral energy distributions (SEDs) of quasars do not depend significantly on bolometric luminosity or redshift \citep[e.g.][]{temple2021}. Thus, the SEDs of $M_{i} < -23$ quasars are representative of quasar SEDs in the full SDSS quasar catalogue. Dust-reddened quasars were removed using the ratio of the median fluxes between $4425-4775$\,\AA\, and $3300-3600$\,\AA\ in the quasar rest frame, requiring the ratio to be $< 0.8$. This eliminated the reddest 20 per cent of objects. Finally, we removed a small number of quasars with low-ionisation BAL troughs. This gives us a final training sample of 2399 quasars.

\subsubsection{Spectra acquisition and pre-processing}

We use publicly available galaxy and quasar spectra from the Sloan Digital Sky Survey (SDSS) seventh data release (DR7) \citep{abazajian2009}, with OH molecule emission subtracted using the PCA prescription of \citet{wild2005}. We mask the $4279$, $5578$, $5894$, $6300$ and $6366$\,\AA\ telluric emission lines and correct spectra for foreground Milky Way dust extinction using the \citet{odonnell1994} dust extinction model and SDSS pipeline E(B-V) values from \citet{schlegel1998}. We shift all SDSS spectra to their rest frames using redshifts from \citet[][]{hewett2010}, and restrict the spectra to a rest frame wavelength range of $3300-5200$\,\AA. We chose this wavelength range as it contains many stellar absorption features, including the 4000\,\AA\ and Balmer breaks, the $\rm H\delta$ and higher order absorption lines, and important quasar features such as the blue continuum and broad $\rm H\beta$ emission. The $3300-5200$\,\AA\ interval also includes the transition from AGN light domination at bluer wavelengths to galaxy light domination at redder wavelengths. To ensure that the full  rest-frame wavelength interval is included in the spectra, we select objects in the redshift range $0.16 \leq z \leq 0.76$. Finally, we normalised the spectra using the mean of the fluxes in the $3300-5200$\,\AA\ region.

\subsection{Generating MFICA components}\label{makecomp}

For the generation of positive components, we use an exponential prior as defined in \citet{Allen2013}. When the positivity needs to be dropped, we use the Laplace prior defined in \citet{Allen2013}.

\subsubsection{Generating galaxy components}

\begin{figure*}
    \centering
    \includegraphics[width=0.8\textwidth]{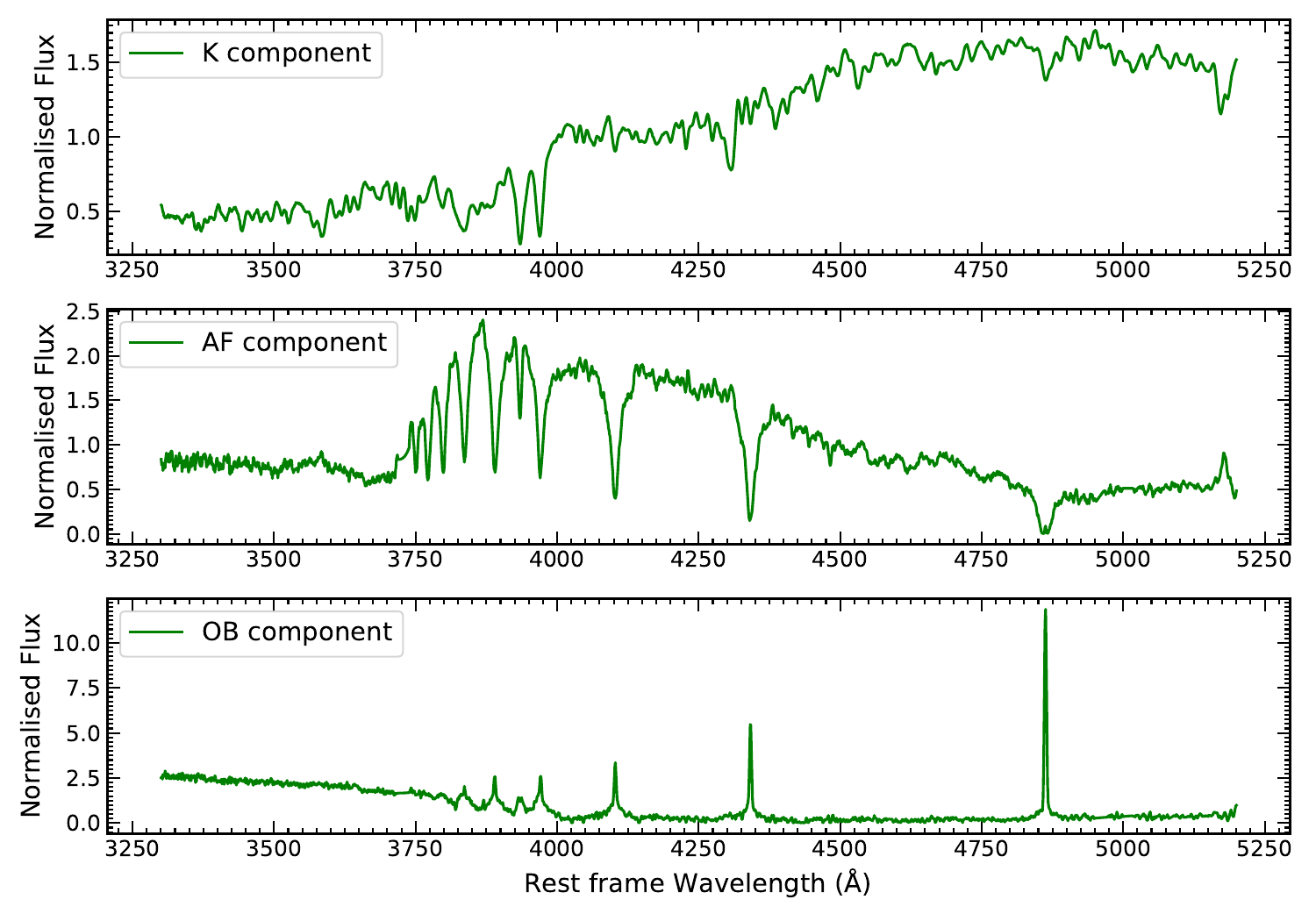}
    \caption{The MFICA galaxy components ($f_\lambda$). From top to bottom: the K component, AF component and OB component.}
    \label{fig:gal_comps_main}
\end{figure*}

Using MFICA, we first derive galaxy components that encode the light emitted by stars and gas in a galaxy - a task shown by \citet[][]{kaban2005}, \citet{lu2006} and \citet{Allen2013} to be feasible with MFICA. In principle, a large number of components can be generated to allow reconstruction of galaxy spectra essentially perfectly. However, this may produce components that are degenerate with features in the quasar components and cause over-fitting of the spectra. We thus limit the number of galaxy components we generate to mitigate this possibility.

Using components representing young, intermediate-age and old stars, we expect to be able to reconstruct the stellar continuum and mean nebular emission of all galaxies to a sufficient extent for our purposes. Additional components are then derived to describe the variations in emission line ratios. The PCA classifications of our training sample galaxies provide prior information on their stellar populations which we can use to simplify the galaxy component creation.

We start by running MFICA on the sample of red galaxies. Red galaxies are composed of old stars with a low variance between spectra. We find that one positive component can reconstruct the spectra of most red galaxies to a level sufficient for the purposes of this paper. Given the similarity of the component to the mean of the red galaxy spectra in the training galaxy sample, we simply use the mean red galaxy spectrum as the first component. This component, shown in the top panel of Fig.~\ref{fig:gal_comps_main}, is labelled the K component as it resembles a K-giant star spectrum.

Next, we apply MFICA to the PSBs in the training sample. The spectra of PSBs are the next simplest, containing intermediate-age and old stars. We run MFICA looking for two positive components, including the previously determined K component. The MFICA analysis returns the component shown in the middle panel of Fig.~\ref{fig:gal_comps_main}, with signatures of A-type stars such as strong Balmer absorption lines. This component is labelled the AF component.

The remaining galaxy classes are composed of mixtures of young, intermediate-age and old stars. We run MFICA on the entire training sample of 3370 galaxies, looking for six positive components, including the previously determined K and AF components. The first of the new components, shown in the bottom panel of Fig.~\ref{fig:gal_comps_main}, adds active star formation and O/B stars. However, it does not take the shape of an O/B type stellar spectrum due to the positivity constraint of our MFICA analysis, which forces the component to infill Balmer absorption and reduce spectral signatures of A and F stars.

\begin{figure*}
    \centering
    \includegraphics[width=\textwidth]{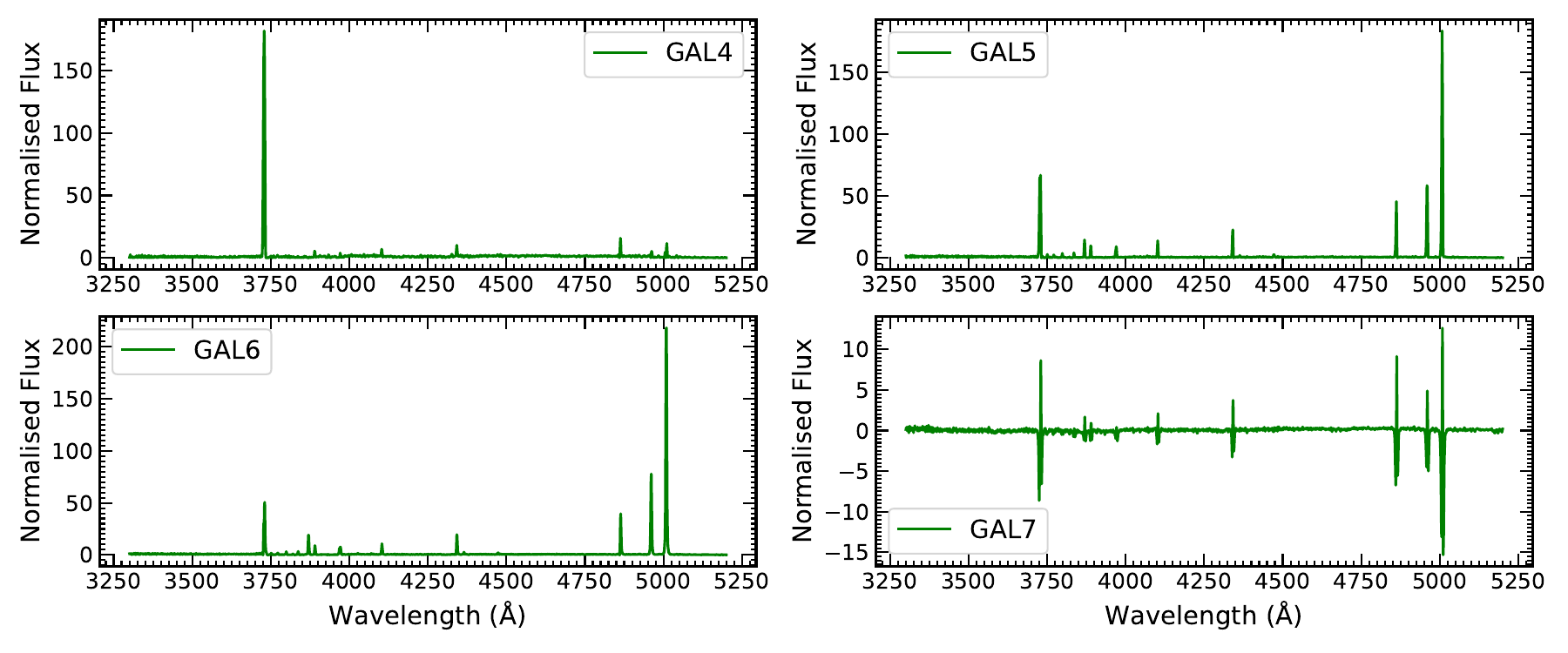}
    \caption{The MFICA emission line components (GAL4-GAL6) and the corrective component (bottom right; GAL7).}
    \label{fig:gal_comps_em}
\end{figure*}

The remaining three components, shown in Fig.~\ref{fig:gal_comps_em}, adjust emission line ratios and are labelled GAL4-6. We generate an additional corrective component, GAL7, to improve emission line reconstructions by altering velocity widths. This component was created by running MFICA on all the galaxy spectra with the contribution of the previous components subtracted, and the positivity constraint dropped.

\subsubsection{Generating quasar components}\label{sec:QSOcomponents}

\begin{figure*}
    \centering
    \includegraphics[width=\textwidth]{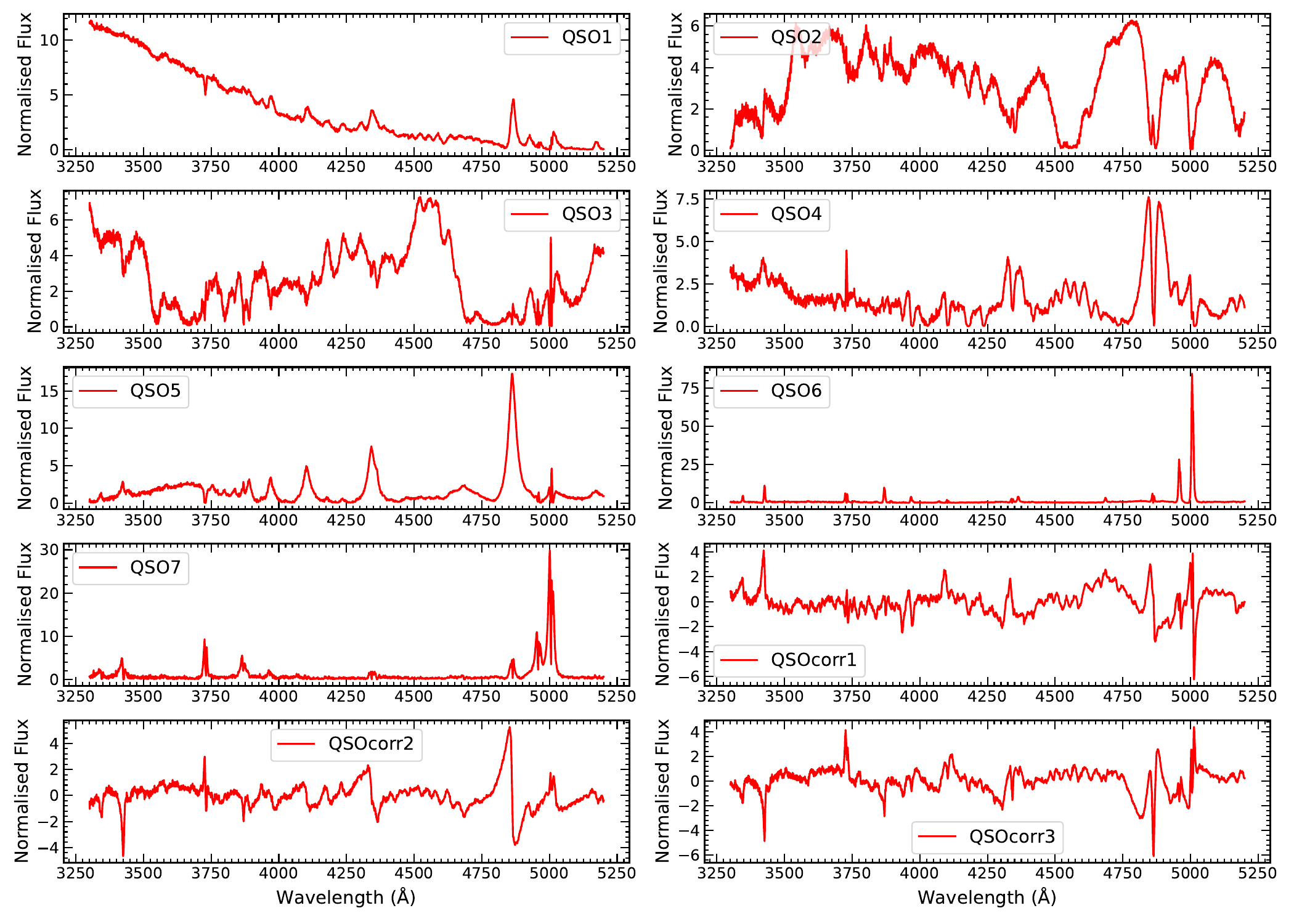}
    \caption{The MFICA quasar components ($f_\lambda$), after subtraction of host-galaxy contamination. The first seven components (QSO1-7) are the main positive components. The remaining three (QSOcorr1-3) are the corrective components generated with no positivity constraint.}
    \label{fig:qso_comps}
\end{figure*}

We leverage the ability to pre-specify the form of components in the MFICA code to generate quasar components statistically independent of MFICA galaxy components. To account for contamination of the quasar spectra by light from the host galaxies, we first run MFICA on the quasar training sample with the K component fixed as the first component. We extract seven positive quasar components free of old stellar population signatures, such as the \caii and \mgi$\lambda5175$ absorption lines. The K component contributes an average of $\sim7$ per cent by light to each quasar spectrum. We subtract the contributions of the K component from each spectrum, giving us training spectra free of old stellar population light. This step impacts the measured stellar mass of the host galaxies, and is slightly sensitive to the input quasar sample and the wavelength range of the components. We test our results with and without this step, and note where this introduces additional uncertainties to our results.

Next, we run MFICA on the K-subtracted quasar spectra with the AF component fixed instead. The average fractional contribution of the AF component to these spectra is $\sim2$ per cent. Despite this small contribution, we subtract the AF contribution from the quasar spectra to minimise the presence of features such as strong Balmer absorption lines. We choose not to repeat the host galaxy subtraction with the OB component, owing to its similarity to an AGN continuum. Based on the average K and AF contributions however, the average OB contribution is expected to be $\sim1$ per cent (see Fig.~\ref{fig:gal_weights}). This contribution is smaller than the error on the quasar components.

Finally, we need to ensure the QSO components are not contaminated by nebular emission from \hii regions, as this could affect our ability to recover the galaxy stellar population, due to the correlation between \hii region nebular emission and stellar continuum shape. To do this, we first remove the continuum of the combined quasar+galaxy spectra using a 61 pixel width median filter, which effectively isolates the narrow emission-lines. We then run a new MFICA decomposition on the median-filtered spectra, fixing components GAL4, 5 and 6 and generating one further component \citep[see, e.g.][]{Allen2013}. This  new component isolates the broader and often asymmetric AGN narrow-line-region lines. We then subtract the GAL4, 5 and 6 contributions from the K- and AF-subtracted quasar spectra and re-run the MFICA to generate galaxy-free QSO components.

Our emission line subtraction routine results in nearly all the \oii\,$\lambda$3727,\,3729 emission being assigned to the MFICA galaxy components, while GAL4, 5 and 6 contribute to the narrowest component of the \oiii\,$\lambda$4959,\,5007 emission. The quasar components reproduce the higher-ionisation narrow emission (e.g. \neiii\,$\lambda$3870 and \nev\,$\lambda$3427), the significantly broader velocity-width contributions to most of the lines and the blue-asymmetric outflow components (particularly for \oiii\,$\lambda$4959,\,5007) that are common in quasar spectra \citep[e.g.][]{mullaney2013}. The average subtracted flux from GAL4, 5 and 6 is similar to the average flux of these components in starburst galaxies.

The seven MFICA quasar components obtained are shown in Fig.~\ref{fig:qso_comps}, labelled QSO1 to QSO7. These components reconstruct the galaxy-subtracted quasar spectra with an average root-mean-squared (rms) error of 7.4 per cent with respect to the quasar continuum. We generate three additional corrective components (with the positive prior dropped) from quasar-component-subtracted quasar spectra to improve the reconstruction of small residual features. The inclusion of these three corrective components reduces the average rms error to 7.3 per cent. 

\section{Examination of galaxy components}\label{val_gal}

\subsection{Validating the galaxy components}

\begin{figure*}
\centering
\includegraphics[width=\textwidth]{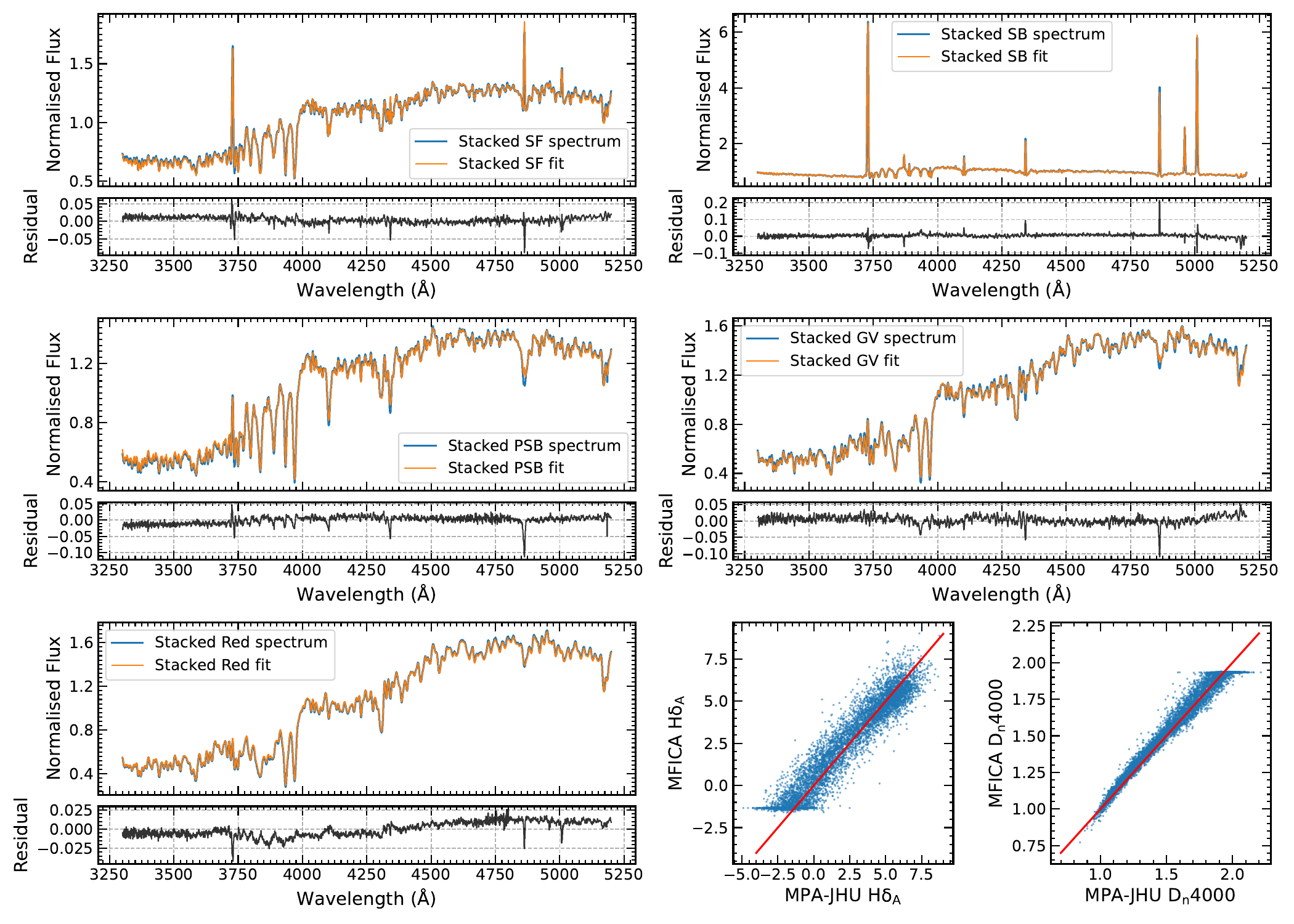}
\caption{Stacked spectra and stacked reconstructions demonstrating the effectiveness of MFICA galaxy reconstructions. Upper panel of each subplot: The stacked (mean) spectra and stacked MFICA galaxy component fits of all spectra of a given galaxy spectral type are plotted in blue and orange respectively. Lower panel of each subplot: Difference (residual) between the galaxy spectra and fitted spectra stacks. The last two panels on the bottom right are comparisons of H$\delta_{A}$ (left) and $\rm D_{n}4000$ (right) measured from MFICA reconstructions on the y-axes, compared to their MPA-JHU catalogue values on  the x-axes. The red line in these plots is the line of equality.}
\label{fig:stacked_recon}
\end{figure*}

We test the effectiveness of the MFICA galaxy components by fitting the parent sample of galaxies with the linear combination of MFICA galaxy components. We obtain the best-fitting linear combination:

\begin{equation}\label{recon}
f_{\lambda, \, gal} = \sum_{i=0}^{7} w_{gal, i} \cdot e_{gal,i}
\end{equation}
where $e_{gal,i}$ are the galaxy components, and $w_{gal,i}$ the weights assigned to each component, by minimising $\chi^{2}$ using the error array provided by SDSS. The $\chi^{2}$ (or weighted least-squares) fitting is handled by the Python package \textsc{lmfit} \citep[][]{newville_2015_11813}, using the default Levenberg-Marquardt least-squares minimiser, which also provides the error on, and covariance between, the components.

To assess the performance of the MFICA components, we fit all the galaxy spectra with the MFICA galaxy components, and then stack (i.e. take the mean of) all galaxy spectra of a given type as well as their best-fit reconstructions. In Fig.~\ref{fig:stacked_recon}, we show the stacked galaxy spectra in orange and the stacked reconstructions in blue. Each panel shows the stacks for a given galaxy type and the residual difference between the stacked spectra and reconstructions. The MFICA galaxy components fit the galaxy spectra remarkably well given the small number of components, achieving a median reduced $\chi^{2}$ of 1.02 and an average rms error of $\sim1.2$ per cent across all the stacks. Small residuals are present around \caii, the Balmer emission lines, and the \oii and \oiii doublets. While residuals could be further reduced by using more components, for this paper we deliberately chose to limit the number of components to avoid unwanted degeneracies or over-fitting when fitting the host galaxies of quasars.

We also check if the MFICA can recover H$\delta_{A}$ and $D_{n}4000$ measurements of the galaxy spectra. $D_{n}4000$ \citep[as defined in][]{balogh1999} increases with the age of a galaxy’s stellar populations \citep{Kauffmann2003agn}. H$\delta_{A}$ \citep[as defined in][]{worthey1997} is largest for star-forming galaxies, or galaxies that have experienced a recent ($\leq 1-2$\,Gyr) burst in star formation \citep{Kauffmann2003agn}.We use the emission-line free PCA components of \citet{Wild2007} to subtract emission lines from the MFICA reconstructions before calculating the H$\delta_{A}$ and $D_{n}4000$ of the MFICA reconstructions. The last two panels in the bottom right of Fig.~\ref{fig:stacked_recon} show the H$\delta_{A}$ and $D_{n}4000$ recovered by MFICA compared to those from the MPA-JHU catalogue \footnote{The hard cuts on $D_{n}4000$ and H$\delta_{A}$ are caused by the use of a single red galaxy template, as opposed to an inherent limitation of MFICA.}. The recovered values all lie close to the line of equality, proving that the MFICA galaxy components capture the properties of galaxies without bias.

\begin{figure*}
\centering
\includegraphics[width=\textwidth]{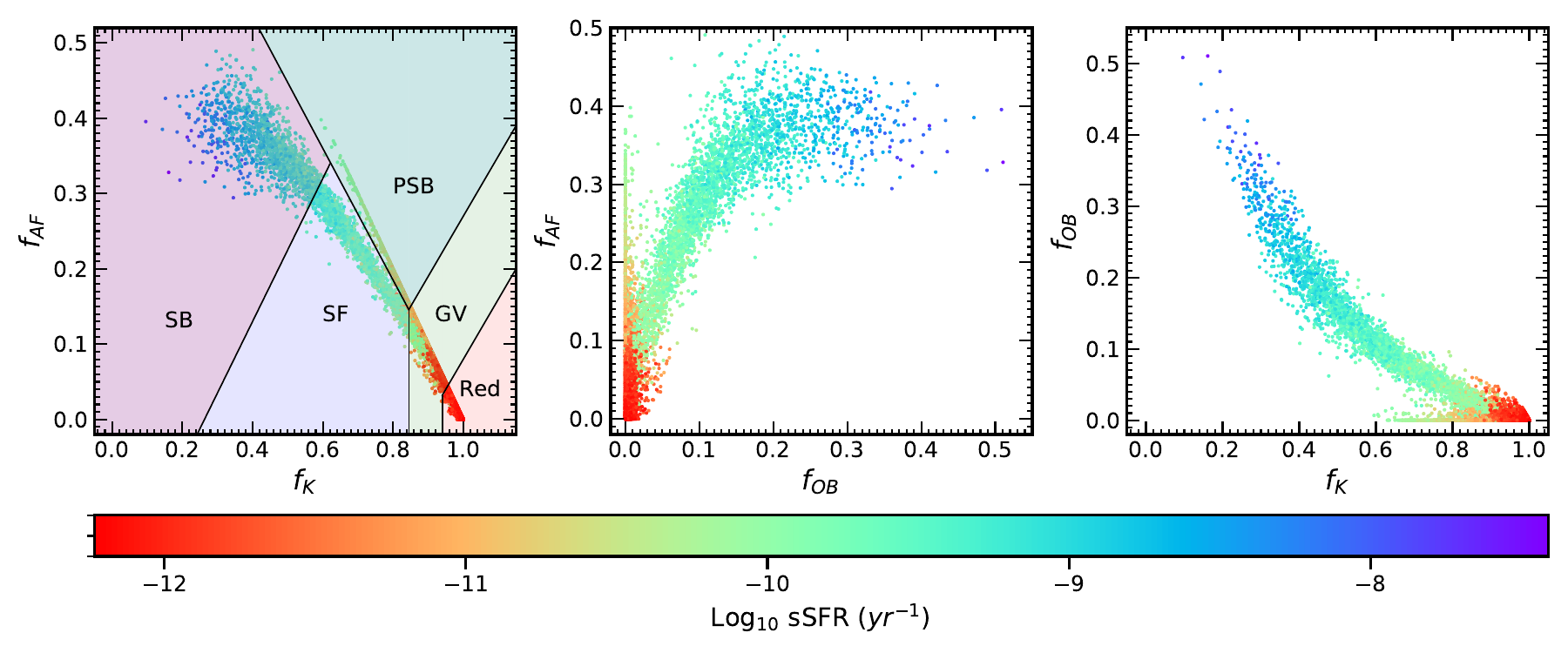}
\caption{Scatter plots of galaxy component fractions, derived from fitting the parent galaxy sample with the MFICA galaxy components. Galaxy component fractions (dots) are colour coded by specific star-formation rate (sSFR) from \citet{Brinchmann2004}. \emph{Left:} K-component vs.  AF-component fraction. The boundaries in this panel demarcate the approximate regions for different galaxy types. These boundaries are used in quasar host galaxy classifications. \emph{Middle:} OB-component vs. AF-component fraction. \emph{Right:} K-component vs. OB-component fraction. In all three plots, the primary axis of variation is the sSFR sequence of normal galaxies, while the post-starburst galaxies form a distinctly separate population in MFICA space.}
\label{fig:gal_weights}
\end{figure*}

\subsection{Galaxy component fractions}\label{gal_comps}

The MFICA galaxy components resemble the spectra of stellar populations of different ages. The weight assigned to these components thus encodes information on the abundance of stellar populations in galaxies. To explore this, we define three quantities - $f_{K}$, $f_{AF}$ and $f_{OB}$ - that represent the fractional contribution of each component to the total flux from these three components integrated across the $3300-5200$\,\AA\ interval. That is, the fraction of the i\textsuperscript{th} galaxy component is defined as:

\begin{equation}\label{comp_frac}
    f_{gal,i} = \frac{\left<w_{gal,i} \cdot e_{gal,i}\right>}{\left<\sum_{i=1}^{3} w_{gal,i} \cdot e_{gal,i}\right>}
\end{equation}
where $e_{gal,i}$ is the flux of the i$\textsuperscript{th}$ galaxy component, $w_{gal,i}$ are the weights defined in Eqn.~\ref{recon}, and $\left< \right>$ represents the mean over the full $3300-5200$\,\AA\ wavelength range. The sum in the denominator represents the sum of the fluxes assigned to the K, AF and OB components only. For example, an $f_{K}$ of 0.8 implies that 80 per cent of the reconstructed galaxy's continuum light comes from the K component.

\begin{figure*}
\centering
\includegraphics[width=\textwidth]{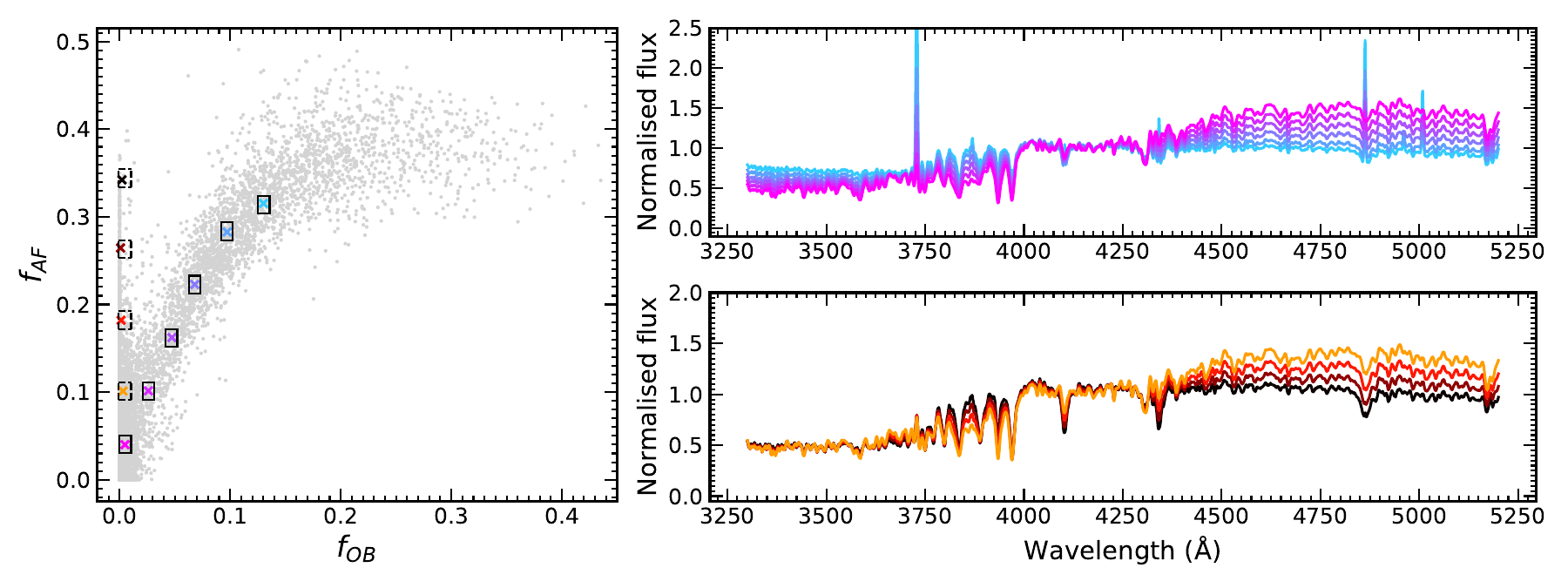}
\caption{Demonstration of how a change in AF or OB component fraction corresponds to a change in shape of a galaxy spectrum. \emph{Left:} AF-component vs. OB-component fraction of the parent galaxy sample. The black boxes demarcate regions from where galaxy spectra were stacked. The solid (dashed) boxes correspond to the stacked spectra in the top (bottom) right panel. \emph{Top right:} Stacked spectra of different regions in the "main-sequence" of galaxies in AF and OB component fractions. For each stack, the corresponding AF and OB component fractions are marked by the cross of the same color. \emph{Bottom right:} Stacked spectra of different regions within the post-starburst branch in $f_{AF}$ and $f_{OB}$ space. All the spectra have been normalised to their mean value between 4100 and 4200\,\AA.}
\label{fig:wt_dem}
\end{figure*}

The component fractions of the parent galaxy sample are plotted in pairs in Fig.~\ref{fig:gal_weights}, with objects colour-coded by their specific star-formation rates (sSFRs)  where the sSFR was measured using H$\alpha$ luminosity, corrected for dust attenuation using the Balmer decrement
\citep[i.e. $i_{class}= 1$ or $2$ in][]{Brinchmann2004}\footnote{For the purposes of this plot only, we excluded galaxies with H$\alpha$ equivalent width $<3$\AA, due to contamination of this measurement for ``retired'' galaxies \citep[][]{cidfernandes2010}.}. In Fig.~\ref{fig:wt_dem}, we focus on the middle panel of Fig.~\ref{fig:gal_weights} and show how the shape of a galaxy spectrum changes as the component fractions change. These plots show that the components do encode the abundance of stellar populations. Objects with low $f_{AF}$ and $f_{OB}$, but $f_{K}$ close to unity are red/quiescent galaxies with low sSFRs, dominated by old stellar populations. On the other hand, $f_{AF}$ and $f_{OB}$ are highest (and $f_{K}$ close to zero) for star-forming and starburst galaxies with abundant young stellar populations and high sSFRs. As $f_{AF}$ and $f_{OB}$ increase, and $f_{K}$ decreases, the stellar populations become younger. As shown in the top right panel of Fig.~\ref{fig:wt_dem}, moving from low $f_{AF}$ and $f_{OB}$ weight (pink) to a high $f_{AF}$ and high $f_{OB}$ weight (cyan) results in the flattening of the galaxy spectra as their 4000\,\AA\ break strengths decrease and emission line strengths increase.

Additionally, for the same $f_{K}$, galaxies which have experienced a recent burst of star formation have a higher $f_{AF}$ and lower $f_{OB}$. This is expected for post-starburst galaxies with a dearth of short-lived ($\sim100$\,Myr lifetime) O/B stars after a rapid burst and subsequent cessation in star formation, but abundant A/F type stars. Moving along this branch of post-starburst galaxies (bottom right panel of Fig.~\ref{fig:wt_dem}, from orange to black), the flattening of the spectral slope is accompanied by an increase in Balmer absorption line strength and swap from a 4000\,\AA\ to a Balmer break.

It is thus clear that galaxies with distinct star-formation histories occupy different regions in MFICA component fraction space, allowing MFICA to be used to select/classify galaxies with distinct SFHs. We define boundaries (shown in the left panel of Fig.~\ref{fig:gal_weights}) visually, noting that the location of these boundaries is somewhat arbitrary, apart from the positioning of the post-starburst galaxies where a strong bimodality is seen. The membership criteria based on these boundaries is as follows:
\begin{enumerate}
    \item starburst: ($f_{AF} < -0.875f_{K} + 0.885$) and ($f_{AF} > 0.95f_{K}- 0.25$)
    \item star-forming: ($f_{AF} < 0.95f_{K} - 0.25$) and ($f_{AF} < -0.875f_{K} + 0.885$) and ($f_{K} < 0.845$)
    \item post-starburst: ($f_{AF} > -0.875f_{K} + 0.885$) and ($f_{AF} > 0.8f_{K} - 0.53$)
    \item green-valley: ($f_{K} > 0.845$) and ($f_{AF} < 0.8f_{K} - 0.53$) and (($f_{K} < 0.94$) or ($f_{AF} > 0.8f_{K} - 0.72$))
    \item quiescent: ($f_{K} > 0.94$) and ($f_{AF} < 0.8f_{K} - 0.72$).
\end{enumerate}

We will use these boundaries and membership criteria to classify inactive galaxies and quasar host galaxies, based on their star-formation histories.

The distribution of the component fractions closely resembles the H$\delta_{A}$-$D_{n}4000$ diagram \citep{Kauffmann2003gal} and the PC1-PC2 diagram \citep{Wild2007}. All three diagrams encode information on the star-formation histories of galaxies, although the MFICA component weights are more directly related to the stellar populations of galaxies.

\section{Validating quasar spectral decomposition with mock spectra}\label{validation}

In this section, we describe how we decompose spectra using the MFICA quasar and galaxy components and validate our approach by constructing and decomposing mock spectra. In Section \ref{decomp}, we describe the decomposition procedure. In Section \ref{mock_create}, we describe the construction of mock spectra, and in Section \ref{mock_decomp}, we assess the quality of the mock decomposition. Finally, in Section \ref{stellar_mass}, we describe our technique to measure quasar host galaxy stellar masses. To determine the efficacy of the decomposition for SDSS spectra, we ask the following questions:

\begin{enumerate}
    \item How well can MFICA decomposition recover the fraction of the total light contributed by the host galaxy ($f_{gal}$)?
    \item How accurately can we recover the galaxy component fractions? In what limits can we accurately classify the host galaxies of quasars?
    \item How does the recovery of component fractions vary as a function of a) S/N of the mock spectra; b) $f_{gal}$ of the mock spectra?
\end{enumerate}

\subsection{Decomposing quasar spectra with MFICA}\label{decomp}

In order to decompose a quasar spectrum into its quasar-only and host galaxy parts, we fit it with a linear combination of the MFICA galaxy and quasar components simultaneously:

\begin{equation}
\centering 
f_{\lambda,\,total} = \left( \sum_{i=0}^{7} w_{gal,i}\cdot e_{gal,i}\right) + \left(\sum_{j=0}^{10} w_{qso,j}\cdot e_{qso,j}\right)
\end{equation}
where $e_{gal,i}$ is one of 7 galaxy components, $e_{qso,j}$ is one of 10 quasar components, $w_{gal,i}$ and $w_{qso,j}$ are the weights assigned to the galaxy and QSO components.

We fit the model to spectra using \textsc{lmfit}, following a three-step procedure:

\begin{enumerate}
   \item First, all galaxy and quasar components, barring the corrective components, are fit together. Each component is assigned an initial weight of 0.1, with a minimum of 0 as the components are positive.

    \item Next, we repeat the full fit with the inclusion of galaxy and quasar corrective components. We assign an initial value of zero to the corrective components and allow the quasar corrective weights to vary within $+/-3\sigma$ of their known distributions, as determined from the quasar training sample. All other components are fixed at their previously obtained values.

    \item In the final step, we drop any components whose weights are less than 0.0001, simplifying the model fit, which in turn allows \textsc{lmfit} to estimate errors for all component weights \citep{newville_2015_11813}.

\textsc{lmfit} returns errors on the component weights ($w_{gal,i}$ and $w_{qso,j}$) along with a covariance matrix \citep{newville_2015_11813}. We use error propagation to obtain errors on the galaxy component fractions ($f_{K}$ and $f_{AF}$) from the errors on the galaxy component weights ($w_{gal,i}$), following Eqn. \ref{comp_frac}.

In Section \ref{sec:quasar_host_galaxy_properties} below, we calculate the percentage of quasar host galaxies in each of the five star formation history categories shown in Fig.~\ref{fig:gal_weights}. To incorporate uncertainties into this calculation, fully accounting for correlations between the components, we resample the component fractions and re-calculate the fraction of quasar host galaxies in each category 10,000 times using the covariance matrices. The mean and standard deviation of the distributions are reported as the final percentages and errors for each star-formation history category.

\end{enumerate}

\subsection{Creating mock spectra}\label{mock_create}

While we assume that a quasar spectrum can be decomposed by fitting it with linear combinations of MFICA components, the success of such an approach is not guaranteed. We cannot, however, test our assumption using real quasar spectra, as properties of their host galaxies are not known a priori. We thus create mock spectra, where input parameters are known, to test the MFICA decomposition.

We adopt the following recipe to create mock spectra that resemble real SDSS DR7 spectra:

\begin{enumerate}

\item For the host galaxy, we construct high S/N composite spectra by taking the mean of all galaxy spectra in each spectral class in our parent sample: star-forming (SF), starburst (SB), post-starburst (PSB), green-valley (GV), and quiescent (red). These composites are not perfectly reconstructed by the MFICA components (see Fig.~\ref{fig:stacked_recon}), and are thus out of MFICA parameter space, ensuring that decomposition of mocks is non-trivial.

\item For the quasar, we use MFICA reconstructions of the  quasar training sample. This ensures the mocks represent the full diversity of spectra seen across the quasar population, and also that the mock quasar spectra are free from host galaxy contamination.

\item We combine random pairs of quasar reconstructions and galaxy composite spectra using a random value (between 0 and 1) for the fraction of light coming from the host galaxy ($f_{gal}$). 

\item We add noise to our mock spectra using an error array from a random SDSS quasar spectrum (MJD $=51817$, plate $=418$, fiber $=566$), scaled to achieve a specific median S/N between $3300-5200$\,\AA. We resample the fluxes of the mock spectra from a normal distribution with $\sigma$ equal to the scaled error array.

\end{enumerate}

\subsection{Results from decomposing mock spectra}\label{mock_decomp}

\begin{figure}
\includegraphics[width=\columnwidth]{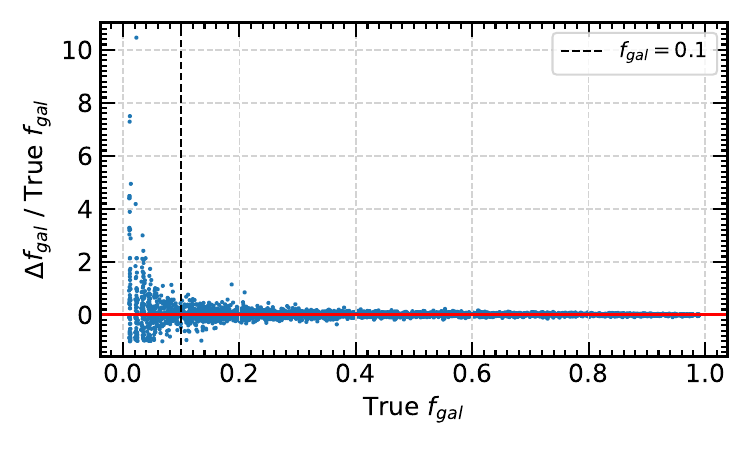}
\caption{Results from mock spectra to demonstrate the fractional accuracy of host galaxy fraction ($f_{gal}$) recovered from the MFICA quasar spectral decomposition, as a function of the true $f_{gal}$. The dashed line represents the lower limit of $f_{gal} = 0.1$ below which the recovered $f_{gal}$ is no longer accurate.}
\label{fig:fgal_mock}
\end{figure}

\begin{figure*}
\includegraphics[scale=0.60]{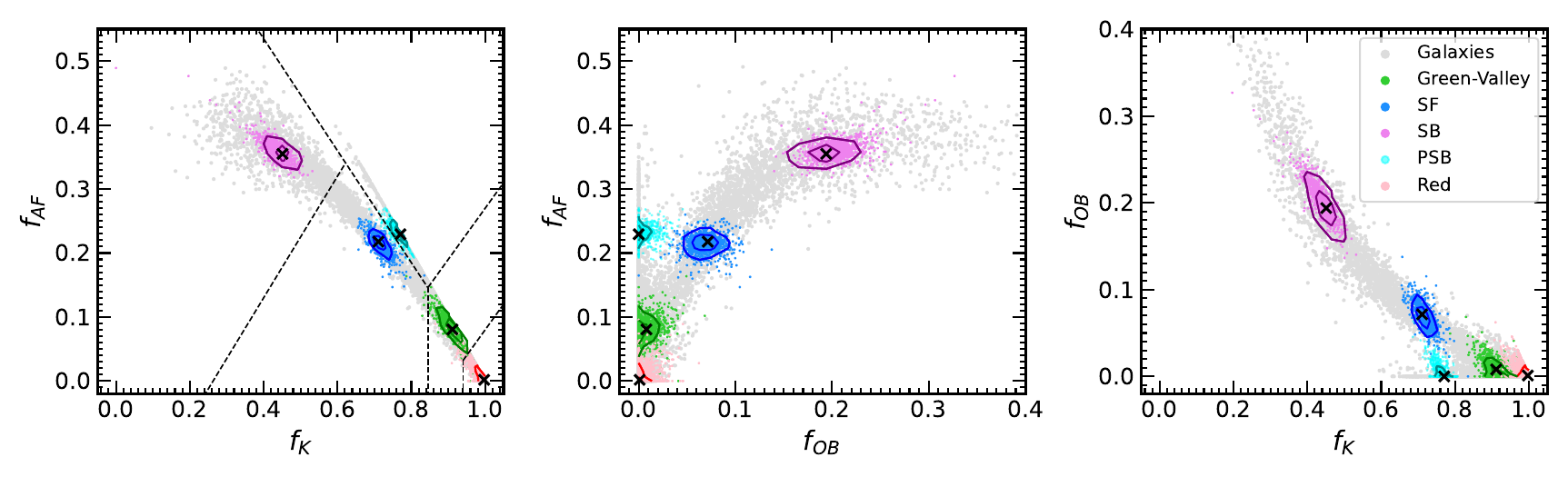}
\caption{Scatter plots of the galaxy component fractions from decomposing mock quasar+galaxy spectra with a median S/N of 10 and $0.2 \leq f_{gal} \leq 0.9$, i.e. in the limits where the decomposition is accurate. Crosses represent the true component fractions and contours represent the component fractions recovered from decomposing the mock quasar+galaxy spectra. The contours represent 68 and 95 per cent confidence intervals. \emph{Left:} K-component vs. AF-component fraction. \emph{Middle:} OB-component vs. AF-component fraction. \emph{Right:} K-component vs. OB-component fraction.}
\label{fig:mock_weight}
\end{figure*}

First, we test the recovery of the mean $f_{gal}$, measured over the full $3300-5200$\,\AA\,interval. For each host galaxy composite, we create 1000 mock spectra, assuming a median S/N ratio of 10 (close to the median of the SDSS DR7 quasar catalogue $\sim10.8$), with $f_{gal}$ ranging from 0 to 1, resulting in a total of 5000 mock spectra. We decompose each mock using the procedure outlined in Section \ref{decomp}. In Fig.~\ref{fig:fgal_mock}, we plot the fractional difference in the recovered $f_{gal}$ compared to the true $f_{gal}$, as a function of the true $f_{gal}$. We find that MFICA can accurately (rms scatter in $\Delta f_{gal}/f_{gal}$ of 0.09) recover $f_{gal}$ for $f_{gal}\geq 0.1$. Below this limit, the recovery of $f_{gal}$ worsens significantly.

Next, we quantify the S/N ratio and $f_{gal}$ limits above which galaxy component recovery is accurate. The dependence of component fractions on $f_{gal}$ and S/N ratio are shown in Appendix \ref{sn_fgal_eff}. We create 2000 mock spectra for each host galaxy composite for this task, adopting a random $f_{gal}$ between 0.1 and 0.9. In Fig.~\ref{fig:mock_weight}, we plot the component fractions recovered after decomposition as coloured dots, with contours representing the 68 and 95 per cent confidence intervals for S/N$=10$ mock spectra. We find that at a median S/N ratio $\geq 10$ and an $f_{gal} \geq 0.2$, the 95 per cent confidence contours for the decomposed galaxy component fractions do not overlap, and the frequency of outliers is minimised. In real data, these outliers are characterised by the similarity between the rising continuum fluxes of the OB component and quasars \citep[e.g.][]{bessiere2017}, a known challenge for previous decomposition techniques \citep[][]{matsuoka2015, wu2018}. Above the S/N $\geq 10$ and $f_{gal} \geq 0.2$ limits, $f_{AF}$ and $f_{K}$ show low mean systematic offsets of 0.36 and 0.42 per cent and are well constrained to within an rms spread of 3-4 per cent with respect to their dynamic ranges. We thus consider MFICA decomposition to be accurate above a S/N ratio of 10 and $f_{gal} \geq 0.2$. 

We also tested the decomposition using real galaxy spectra rather than high S/N composites. We find that we are able to recover the demographics of the input mock galaxies accurately.

\subsection{Stellar mass for quasar host galaxies}\label{stellar_mass}

In what follows, we compare our results to a sample of inactive galaxies matched in galaxy stellar mass, which requires an estimate of the stellar mass of the quasar host galaxies. The standard method to estimate the stellar mass of a galaxy is to fit template stellar population models to observed spectra or photometry, to obtain a mass-to-light ratio which is then scaled to the galaxy's luminosity. We tried this standard approach of least squares fitting the reconstructed host galaxy continuum spectra with the ``stochastic burst" library of \citet{bruzual2003} stellar population models, masking the primary nebular emission lines, following the philosophy of \citet{Kauffmann2003gal} and \citet{salim2005}. We tested this method's efficacy for our work by fitting spectra of MPA-JHU galaxies and comparing the resultant stellar masses to the fibre stellar masses in the MPA-JHU catalogue. The MPA-JHU stellar masses were estimated from SDSS fibre aperture photometry using the same standard model fitting approach\footnote{Available from \url{https://wwwmpa.mpa-garching.mpg.de/SDSS/DR7/Data/stellarmass.html}}. We find that the 5200\,\AA\ red wavelength limit used for our components is insufficient to fully constrain the galaxy dust attenuation. This leads to a small ($\sim0.1$dex) overestimate in the stellar masses compared to the MPA-JHU fibre stellar masses, as the model fits tended towards the median of the assumed dust attenuation prior.

Therefore, to better use natural prior correlations between dust attenuation, stellar populations and stellar mass in galaxies, we instead trained a random forest regressor (from \textsc{scikit-learn}, with 100 estimators) on fibre stellar masses of SDSS DR7 spectroscopically classified galaxies in the MPA-JHU catalogues. We normalised these photometric stellar masses by the luminosity of the galaxy spectrum measured over the $3300-5200$\,\AA\ wavelength range to get a mass-to-light (M/L) ratio. The random forest regressor was trained on $f_{AF}$ and $f_{K}$ (obtained from reconstructing galaxy spectra with MFICA) as inputs, and asked to predict M/L. Re-normalising the resulting M/L by the measured galaxy spectrum luminosity gives the final galaxy stellar mass, normalised to the fibre aperture photometry \footnote{We note that the SDSS 'fibre aperture' photometry does not have the same normalisation as the SDSS spectra for galaxies, and our method accounts for this difference.}. Random forests are extremely capable at learning smooth non-linear relationships between variables, as is the case between $f_{AF}$-$f_{K}$ and M/L.

We selected the training and test galaxies for the random forest to have  $0.16 \leq z \leq 0.76$, median S/N$>8$, $i_{class} \leq 2$ \citep{Brinchmann2004} and secure redshifts ($z_{warn} = 0$). A small number of objects were removed with measured $D_n(4000)<0$. This results in 109,180 galaxies, which were randomly split into $70\%$ training and $30\%$ test samples. The stellar masses predicted for the test galaxies have negligible systematic offset compared to the MPA-JHU fibre stellar masses, and an rms scatter of 0.084\,dex, with no systematic deviations for different galaxy spectral types (as defined in Section \ref{gal_comps}. We also verified that different galaxy training sets did not result in different stellar masses for the quasar hosts.

At the redshifts applicable to this work ($z>0.16$), the difference between galaxy stellar masses measured from SDSS fibre and model magnitudes in the MPA-JHU catalogue is 0.5\,dex, independent of redshift. This is because at these redshifts the loss of light is dominated by atmospheric seeing, rather than the fraction of galaxy covered by the physical extent of the fibre. When comparing to other work, we add a constant 0.5\,dex aperture correction to our stellar masses to estimate the total host galaxy mass. Aperture corrections depend on the size and structure of the galaxy, and by using the constant 0.5\,dex correction, we are implicitly assuming the inactive control sample has a similar galactic size distribution to the quasar host galaxies, which may not be entirely correct \citep{Li2021}. However, this aperture correction is not required for our analysis or comparison with the inactive control sample.

\section{Results}\label{results}

\begin{figure*}
    \centering
    \includegraphics[width=\textwidth]{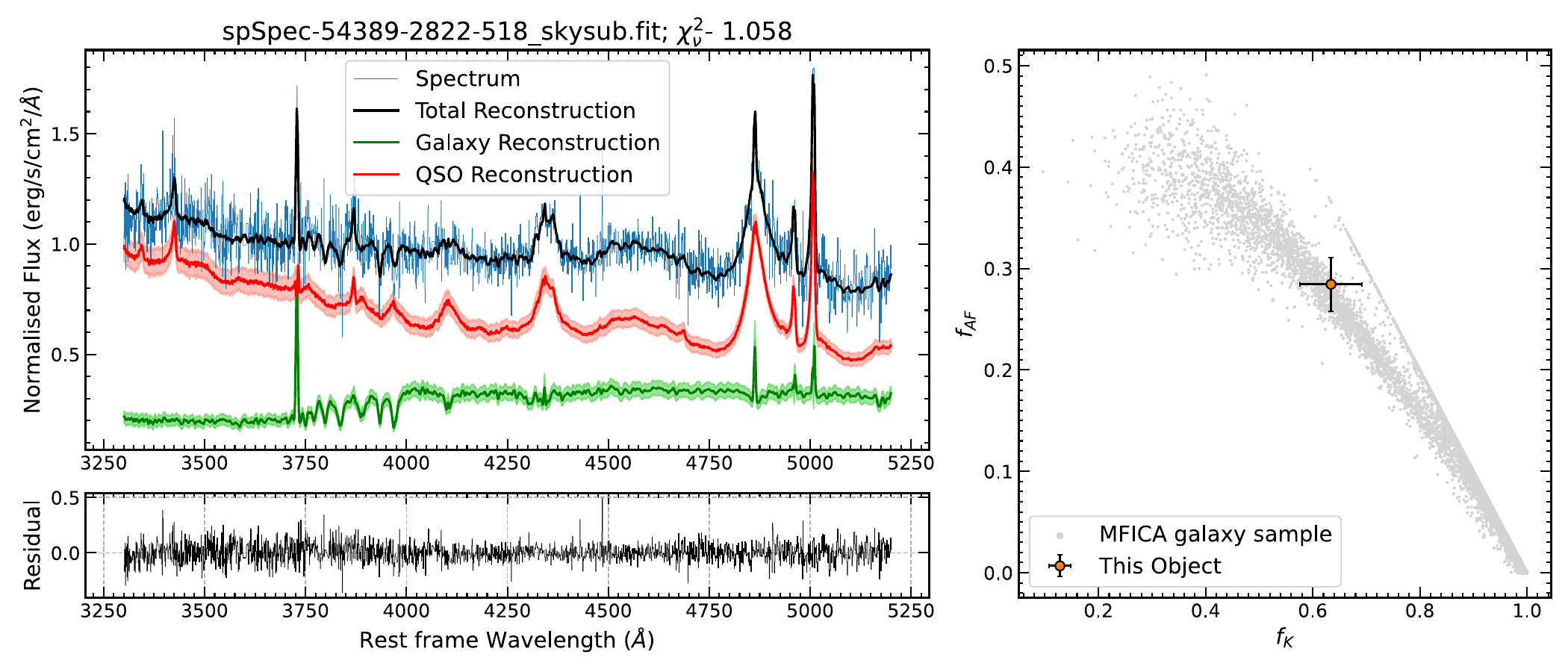}
    \caption{The left panel shows an example of the decomposition of a quasar spectrum. \emph{Top:} The quasar spectrum (blue), best-fit MFICA model (black), reconstructed host galaxy (green) and reconstructed quasar (red). \emph{Bottom:} The difference between the quasar spectrum and the best fit model. The right panel shows the location of this quasar host galaxy (orange dot) in the K-component vs. AF-component fraction space. The error bars are determined from  the errors on the weights from \textsc{lmfit} (see Section \ref{decomp}). The grey dots represent the distribution of the parent galaxy sample.}
    \label{fig:decomp_qso_example}
\end{figure*}

In this section, we apply our MFICA decomposition code to study the stellar populations of quasar host galaxies. We select quasars from the SDSS DR7 quasar catalogue \citep{schneider2010} in the redshift interval $0.16 \leq z \leq 0.76$. For determining the stellar populations of our quasar hosts, we limit this initial sample of 19,358 quasars to those with an $f_{gal} \geq 0.2$, a median S/N ratio $\geq 10$ (see Section \ref{mock_decomp}), and  $\chi^{2}_{\nu} \geq 2.0$ which results in 3376 quasars. This sample of reliably decomposed quasars has a bolometric luminosity range of $44.7 \leq \rm log_{10}(L_{bol}/\rm erg s^{-1}) \leq 47.0$. 

The left hand panel of Fig.~\ref{fig:decomp_qso_example} shows an example of a decomposed quasar spectrum with the reconstructed host galaxy and AGN spectra. There are no significant features in the residuals, and this is typical of the full sample. The majority of quasar spectra are well fit with a median reduced $\chi^{2}$ ($\chi^{2}_{\nu}$) of 1.07. The right hand panel of Fig.~\ref{fig:decomp_qso_example} shows the recovered component fractions of the quasar host galaxy, with error bars calculated as described in Section \ref{decomp}.

\begin{figure}
    \centering
    \includegraphics[width=\columnwidth]{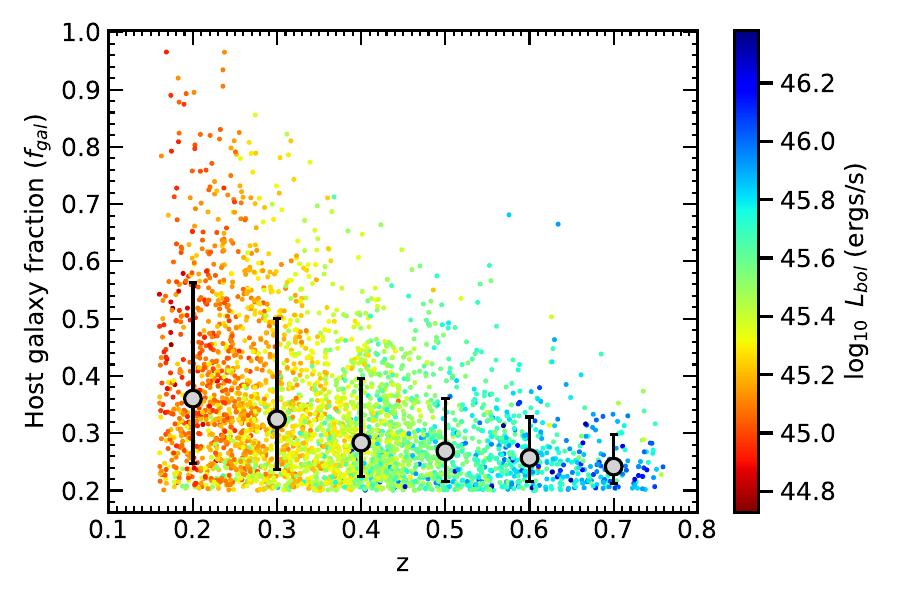}
    \caption{Variation in host galaxy fraction as a function of redshift for quasars with $f_{gal} \geq 0.2$, median $S/N \geq 10$ and $\chi^{2}_{\nu} < 2.0$. The colours correspond to the bolometric luminosity of the quasars \citep{shen2011}. The grey filled circles represent the median $f_{gal}$ in redshift bins of width 0.1, and the bars represent the 16th and 84th percentiles.}
    \label{fig:fgal_dist}
\end{figure}

Fig.~\ref{fig:fgal_dist} shows the distribution of host galaxy fractions ($f_{gal}$) of our sample of quasars, where $f_{gal}$ is measured over the full $3300-5200$\,\AA\ interval. Quasars are colour-coded by their bolometric luminosities ($L_{bol}$) as measured by \citet{shen2011}. The typical $f_{gal}$ decreases with increasing redshift and $L_{bol}$. This trend may be explained by invoking equation (7) of \citet{temple2021}, which relates the galaxy and quasar luminosities to host galaxy fractions. $L_{bol}$ increases with increasing redshift as the SDSS quasar catalogue is flux-limited, including only those objects with $m_{i} < 19.1$ \citep[][]{schneider2010}, which at the highest redshifts correspond to the highest luminosities. However, per the SDSS-calibrated model of \citet{temple2021} the luminosity of galaxies rises slower than the rise in quasar luminosity, resulting in a decrease in $f_{gal}$ with redshift. 

\subsection{Quasar host stellar masses and control sample selection}

\begin{figure*}
\centering
\includegraphics[width=0.75\linewidth]{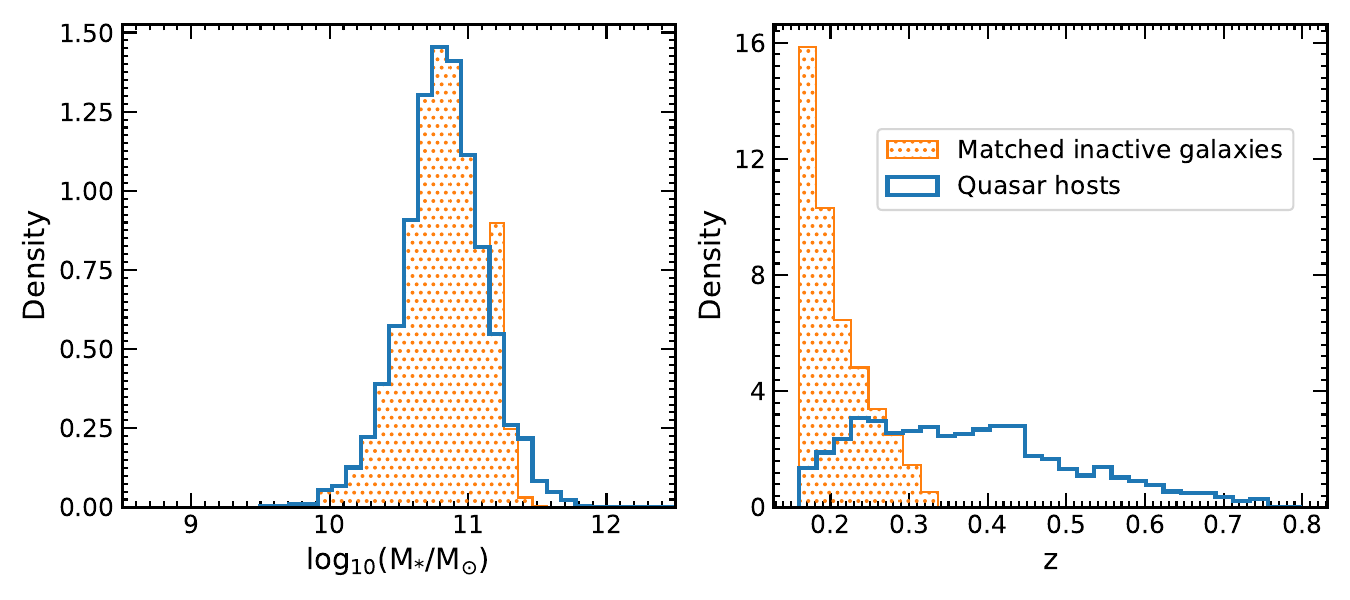}
\caption{\emph{Left:} The normalised histograms of the stellar masses for the quasar host galaxies (solid blue) and the control sample (dotted orange), measured in the 3" SDSS fibre. \emph{Right:} The redshift distribution of the quasar host galaxies (solid blue) and the control sample (dotted orange).}
\label{fig:mass_z_comp}
\end{figure*}

The left panel of Fig. \ref{fig:mass_z_comp} shows the distribution of estimated fibre stellar masses for the quasar host galaxies. Quasars (with $f_{gal} \geq 0.2$) are preferentially hosted by massive galaxies with a median fibre stellar mass (i.e. based on light collected in the SDSS 3" fibre) of $\rm 10^{10.8}\ M_{\sun}$ and an rms deviation from median of 0.3\,dex. This increases to a total stellar mass of $\sim10^{11.3}\ M_{\sun}$ after accounting for a fixed 0.5\,dex aperture correction (Section \ref{stellar_mass}). 

The stellar populations of galaxies depend on stellar mass: galaxies with higher stellar masses have older stellar populations on average. Previously derived differences between the stellar populations of quasar hosts and inactive galaxies may be due to differences in their stellar masses \citep{Trump2013}. We control for the effect of stellar mass when comparing quasar host and inactive galaxies by constructing a stellar-mass matched control sample of galaxies. For each quasar host, we chose two galaxies from the MPA-JHU catalogue with the closest fibre stellar mass. The distribution of fibre stellar masses and redshifts for the 3376 quasars and 6752 control galaxies is shown in Fig. \ref{fig:mass_z_comp}. At the high mass end ($\sim10^{11.5} M_{\sun}$), the fibre stellar mass distributions for the quasar hosts and the control sample do not match exactly. This mismatch is attributable to the larger uncertainties on the stellar masses of the quasar host galaxies which, combined with the steep fall-off in the high mass end of the galaxy stellar mass function, leads to scatter beyond the high mass limit of the control galaxy sample. The number of objects without exact matches is minor, and this effect will not significantly impact the results.

\subsection{Quasar host-galaxy properties}\label{sec:quasar_host_galaxy_properties}

\begin{figure*}
\centering
\includegraphics[width=1.5\columnwidth]{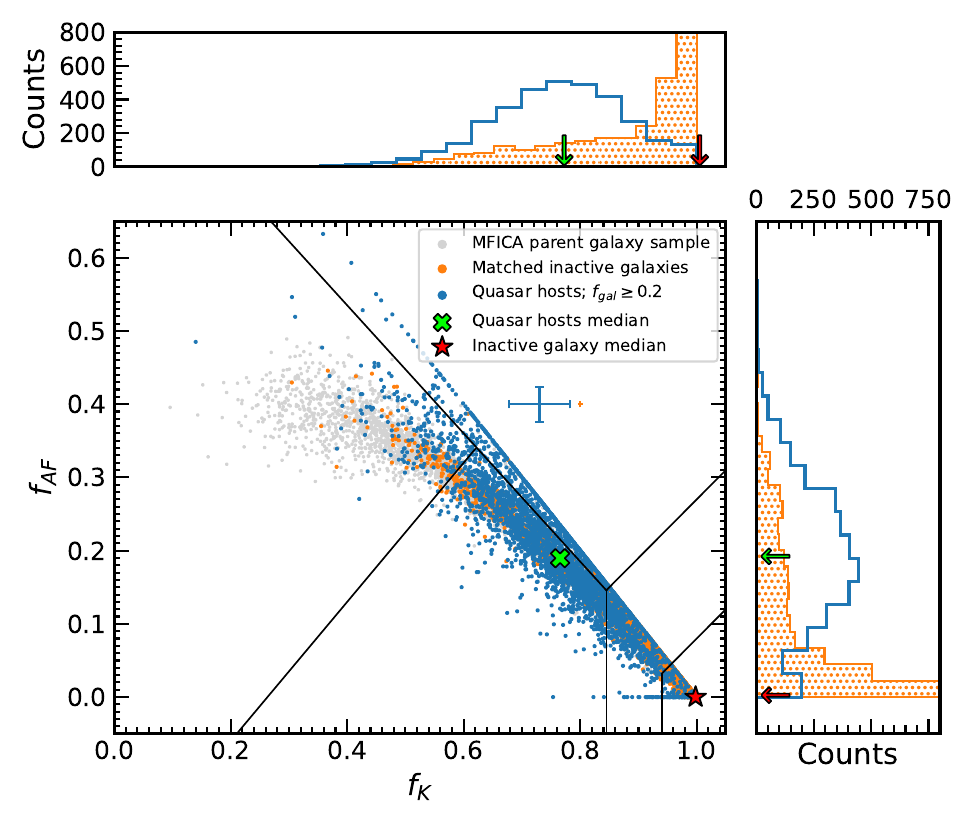}
\caption{K-component fraction vs. AF-component fraction for quasar host galaxies with $f_{gal} \geq 0.2$ (blue dots), and stellar-mass matched control galaxies (orange dots). The light-grey dots represent the distribution of the MFICA parent galaxy sample (for visual-aid only). The green-cross is the median of the weights of all quasar host galaxies, and the red star is the median of the weights of all control galaxies. The error bars represent the mean of the component fraction errors for the decomposed quasar host galaxies (blue) and the control sample (orange). \emph{Right and top:} Projected histograms of AF and K-component fractions for quasar host galaxies (blue) and the control sample (orange). The control sample bins with highest K-component fraction and lowest AF-component fraction have been clipped at 800 counts - there are an additional 3740 galaxies in the final AF-component fraction bin, and 3929 in the final K-component fraction bin.}
\label{fig:cont_vs_qso_sp} 
\end{figure*}

\begin{figure*}
    \centering
    \includegraphics[width=\linewidth]{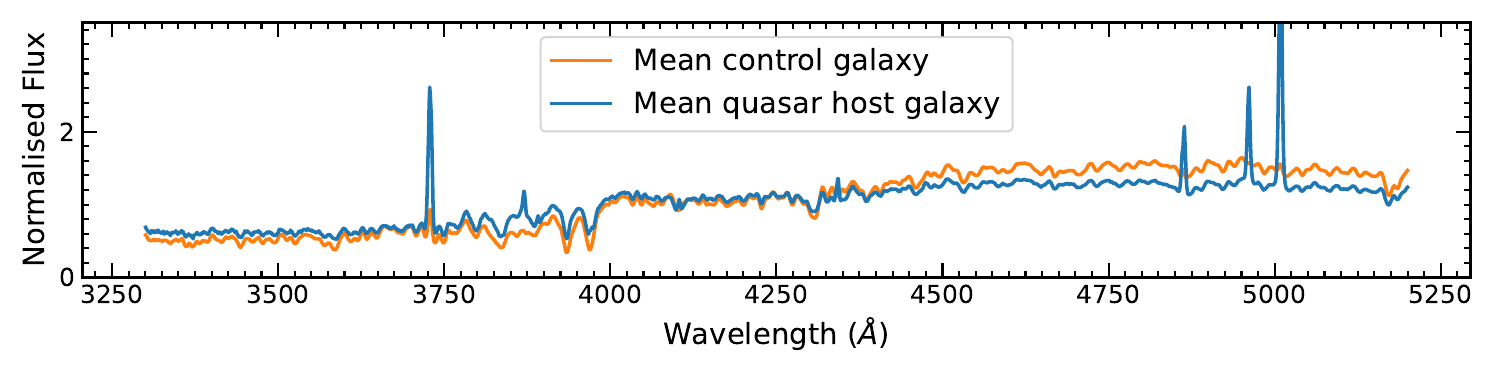}
    \caption{Comparison between the stack (mean) of all quasar host galaxy reconstructions (blue) and the stack of all control galaxy reconstructions (orange). Both stacks are normalised to mean flux of unity. Note that due to the way our MFICA components are constructed, the narrow emission lines in the quasar host spectrum may come from a range of sources that are not only \hii regions.}
    \label{fig:mean_qso_cntrl_comp}
\end{figure*}

In Fig. \ref{fig:cont_vs_qso_sp}, we compare the AF and K-component fractions ($f_{AF}$ and $f_{K}$ respectively) for the quasar hosts and the control sample. Our sample of quasar host galaxies have a wide range of $f_{AF}$ and $f_{K}$. The quasar hosts, however, have generally higher $f_{AF}$ and lower $f_{K}$ than the control sample, implying that quasar host galaxies have younger stellar populations than stellar-mass-matched control galaxies. The younger stellar populations of quasar hosts can also be inferred from the median $f_{K}$ and $f_{AF}$. The median quasar host is a star-forming galaxy, and the median control galaxy is quiescent. The younger stellar populations of the quasar host galaxies can also be seen in Fig. \ref{fig:mean_qso_cntrl_comp}, which compares the stack (mean) of all the reconstructed quasar host galaxy spectra, and the stack of all the reconstructed control galaxies. The stacked quasar host spectrum has a smaller $D_{n}4000$, indicative of a young stellar population and active star formation.

\begin{table}
\centering
\caption{Statistics of quasar hosts with reliable decompositions ($f_{gal} \geq 0.2$) and control sample galaxies. From left to right: (1) The galaxy type based on the AF and K-component fractions. (2) $\rm F_{quasar}$ is the per cent of all quasars that are of a given host galaxy type. (3) $\rm F_{control}$ is the per cent of all control galaxies that are of a given galaxy type.}
\label{tab:qso_dist}
\begin{tabular}{lcc}
\hline
Galaxy & $\rm F_{quasar}$ & $\rm F_{control}$\\
type & $(\text{per cent})$ & $(\text{per cent})$\\
\hline
SB & $6.4\pm0.3$ & $1.59\pm0.03$\\
SF & $46.5\pm0.7$ & $12.22\pm0.06$\\
PSB & $24.9\pm0.6$ & $0.90\pm0.03$\\
GV & $17.5\pm0.4$ & $11.37\pm0.11$\\
Red & $4.6\pm0.2$ & $73.92\pm0.09$\\
\hline
\end{tabular}
\end{table}

\begin{figure}
\centering
\includegraphics[width=\columnwidth]{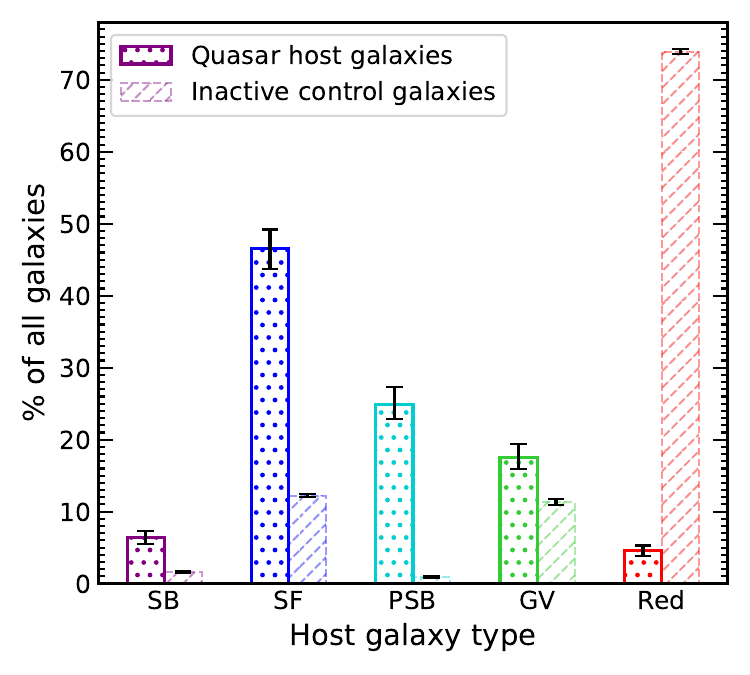}
\caption{Percentage of different types of galaxies in the decomposed quasar host galaxy sample (dot-filled bars) and the control sample (line-filled bars). The height of each bar is the per cent of all quasar host galaxies or control galaxies with a given galaxy type. The error bars represent the maximum and minimum possible per cent of objects of a given galaxy type, based on 10,000 re-samples of the component fractions using the covariance matrix (see Section \ref{decomp}).}
\label{fig:quasar_bar}
\end{figure}

Fig.~\ref{fig:quasar_bar} shows the percentage of different galaxy types in the quasar host and control samples. The height of each bar corresponds to the percentage of either the ($f_{gal} \geq 0.2$) quasar sample or the control sample classified as a particular galaxy type (see the left panel of Fig.~\ref{fig:gal_weights}). The dot-filled bars represent the percentages for the quasar sample, while the line-filled bars represent the percentages for the control sample. The information in this plot is also summarised in Table \ref{tab:qso_dist}. The error bars in Fig.~\ref{fig:quasar_bar} represent the maximum and minimum percentage for each host galaxy type, while Table \ref{tab:qso_dist} shows the true errors in the percentages.

We find that the biggest class of quasar host galaxies are star-forming galaxies ($\sim$47 per cent; $\sim$53 per cent when including starburst galaxies), with a significant number of post-starburst ($\sim$25 per cent) and green-valley hosts ($\sim$ 18 per cent). On the other hand, the vast majority ($\sim$73 per cent) of our stellar mass matched control galaxies are red/quiescent, $\sim$12 per cent are star-forming galaxies, $\sim$2 per cent are starburst, $\sim$11 per cent are green-valley galaxies and $<$1 per cent are post-starburst galaxies. These results imply that the stellar populations and recent SFHs of quasar host galaxies and control galaxies are distinct. 

Apart from a higher fraction of star-forming quasar hosts, we also see that quasar hosts include a larger fraction of post-starburst galaxies than the control sample. While $\sim$25 per cent of all quasar hosts are post-starburst galaxies, $<$1 per cent of our control sample are post-starburst galaxies, which is consistent with previous results \citep[][]{Wild2016, rowlands2018}. Our analysis shows that post-starburst galaxies are $28\pm1$ times more common amongst quasars than in the control galaxies of the same stellar mass. 

The excess of post-starburst galaxies in quasar hosts is substantially driven by the large number of quiescent control galaxies. Therefore, we can alternatively represent the excess of post-starbursts by normalising by the total number of star-forming galaxies. For the same number of star-forming galaxies, it is 7.3 $\pm$ 0.4 times more likely for a quasar to have a post-starburst host than it is for an inactive galaxy to be a post-starburst galaxy.

\subsection{Redshift evolution and stellar mass uncertainties}

The comparatively larger fraction of quiescent galaxies in the control sample could be dictated by their lower redshifts, as the fraction of quiescent galaxies increases over time \citep{noeske2007, illbert2015}. On the other hand, due to cosmic down-sizing, the fraction of massive quiescent galaxies does not change substantially over the redshift ranges probed here \citep{muzzin2013, illbert2013, rowlands2018, mcleod2021, weaver2023}. We are unable to control for redshift, as the redshift distributions of SDSS quasars and galaxies are entirely different \citep{strauss2002, richards2002}. Therefore, we checked if redshift differences drive our results by restricting the quasar sample to $z < 0.3$, close to the maximum redshift of the control sample. The fraction of star-forming quasar host galaxies ($47.9 \pm 1.1$ per cent at $z < 0.3$) does not change within errors, whereas the fraction of quiescent host galaxies rises slightly from $4.6 \pm 0.2$ to $6.2 \pm 0.4$ per cent. Quiescent galaxies, however, still dominate the control galaxy sample even as their fraction drops from $73.92 \pm 0.09$ to $64.9 \pm 0.2$ per cent. This drop occurs due to the lower stellar masses of the quasar hosts (and thus control galaxies) at $z < 0.3$, owing to lower luminosities and consequently (at fixed Eddington ratio) lower black hole masses. These results show that the most important variable is stellar mass, and our conclusion that quasar host galaxies have younger stellar populations than the control galaxies holds firm.

Similarly, the post-starburst fraction in ordinary galaxies increases with redshift, although not particularly strongly for the highest mass post-starburst galaxies studied here \citep[][]{Wild2016,rowlands2018}. For our $z < 0.3$ quasar subset, the post-starburst fraction decreases from $24.9 \pm 0.6$ per cent to $16.7 \pm 0.9$ per cent, while the fraction of post-starbursts in the control sample remains the same within errors. This result implies a substantial excess ($\sim$18x) of post-starburst quasar host galaxies compared to the control sample. Future datasets should allow both mass and redshift matching of active and inactive galaxies, which will be an important next step.

The stellar masses of the quasar host galaxies are slightly sensitive to the exact prescription for removing the galaxy host contamination from the MFICA quasar templates (Section \ref{sec:QSOcomponents}). This does not affect the estimated stellar populations of the quasar hosts, but due to the strong correlation between stellar mass and galaxy stellar populations, it does affect the inferred stellar populations of the inactive control sample. To investigate the impact of this on our conclusions we repeated the analysis with new quasar components which (a) did not remove the galaxy light, or (b) were created from the very brightest quasars where contamination from host light is naturally minimised. We find that the stellar masses of the quasar hosts changes by less than 0.2\,dex in both cases. However, this small change alters the relative proportion of star-forming, green-valley and quiescent stellar populations between the quasars and inactive control sample. We therefore caution that, while the qualitative result holds that the quasar hosts are more star-forming than a mass-matched inactive galaxy sample, the exact amount by which they differ depends in detail on the precise stellar masses of the quasar hosts. On the other hand, the relative fraction of post-starburst galaxies between quasar hosts and inactive control galaxies does not significantly change and is therefore quantitatively robust. Further improvements on measuring the stellar masses of the quasars could come from an extended wavelength range of upcoming galaxy surveys, or combined photometric and spectroscopic decomposition techniques.

\section{Discussion}\label{discussion}
In this section we discuss our results in the context of previous AGN host galaxy studies and the implications for our understanding of the co-evolution of AGN and massive galaxies. 

\subsection{Stellar masses of quasars}

Figure \ref{fig:mass_z_comp} shows that our quasars are hosted by massive galaxies with median fibre stellar mass of $10^{10.8}$ $M_{\sun}$ and aperture corrected total stellar masses of $\sim10^{11.3}$ $M_{\sun}$. It is well established that AGNs generally have massive host galaxies \citep[$M_{*} > 10^{10} M_{\sun}$; ][]{Kauffmann2003agn, Silverman2009, Trump2013}. While physical explanations for this have been proposed \citep[e.g.][]{wang2008, matsuoka2015}, in our sample the preference for massive hosts may be mostly explained by the flux-limited nature of the SDSS quasar catalogue \citep[][]{schulze2011, aird2012}. Lower Eddington ratio quasars are more common than high Eddington ratio quasars \citep[e.g.][]{kelly2013, temple2021}, and low Eddington ratio objects will satisfy the $M_{i} < -22$ inclusion criterion only if their black hole masses are sufficiently large. Given the $M_{\rm BH}-M_{*}$ relation \citep[e.g.][]{reines2015}, this corresponds to large stellar masses \citep[][]{aird2012}. Additionally, quasar host galaxies with lower stellar masses will have lower $f_{gal}$. Such objects may fall below the $f_{gal}$ threshold for a reliable decomposition. 

Independent verification of the high stellar masses of quasar host galaxies comes from clustering studies of SDSS quasars \citep[][]{richardson2012, shen2013}. These studies quote a characteristic halo mass of $\sim 10^{12}-10^{13}$ $M_{\sun}$ that, assuming a halo mass-to-stellar mass function \citep[e.g.][]{behroozi2019}, implies stellar masses in the range $10^{10}$ to $10^{11}{\rm M}_{\sun}$.

Our quasar host galaxy stellar masses are, however, higher than those reported for SDSS Reverberation Mapping (SDSS-RM) quasar samples \citep[once the difference between the 2" BOSS and 3" SDSS fibre apertures is accounted for;][]{matsuoka2014, matsuoka2015, Yue2018}, and those reported for the SDSS quasar catalogue by \citet{Li2021}. The difference compared to the SDSS-RM sample may partly be due to larger black hole masses in our sample \citep{shen2011}, leading to higher stellar masses via the $M_{\rm BH}-M_{*}$ relation. Compared to other decomposition techniques, MFICA may also report higher stellar masses due to the (correct) initial subtraction of host galaxy light from the MFICA quasar components, leaving more host light available to be assigned to the galaxies. There will clearly also be significant differences depending on assumed aperture corrections. Obtaining accurate stellar masses of quasar host galaxies is clearly an important, yet challenging, problem that demands further attention. However, host stellar masses are not the main focus of this work so we leave further examination of this problem to future work.

\subsection{Star-forming host galaxies of quasars}

Figs.~\ref{fig:cont_vs_qso_sp} and \ref{fig:quasar_bar} show that the largest fraction of quasars are hosted by averagely star-forming, but not starburst, galaxies. This result agrees with the overwhelming majority of studies that find that low-redshift ($z \lesssim 1$) optically-selected quasars are hosted by star-forming galaxies \citep[][]{matsuoka2014, Yue2018, Rosario2013, Jahnke2007, Zhuang2022, Trump2013, bettoni2015, Li2021, stanley2017, xie2021, jahnke2004, Ren2024}. These studies use a wide variety of indicators of active star formation, including IR luminosities \citep[e.g.][]{Rosario2013, stanley2017}, optical colours \citep[e.g.][]{matsuoka2014}, emission line SFRs \citep[e.g.][]{Zhuang2022}, and $D_{n}4000$-based stellar ages \citep[][]{Ren2024}. Thus, the predominantly star-forming nature of quasar host galaxies is robustly established. The new result in this paper is the ability to construct a stellar-mass matched control sample, which demonstrates how unusual the star-forming nature of quasar host galaxies actually is, given their stellar masses. We can conclude that powerful quasars do not simply represent a random sub-sample of all galaxies. This in turn suggests that the presence of the AGN and young stellar population are directly or indirectly causally linked.

Previous studies have suggested that stellar populations of AGNs become younger as AGN luminosity increases \citep[][]{Kauffmann2003agn, vdb2006, Silverman2009, Trump2013}. This is analogous to a correlation between AGN luminosity and SFR \citep{Mullaney2012, aird2019}, albeit the evidence for such a correlation has been mixed \citep{rosario2012, stanley2015, mountrichas2022sfr}. To investigate this further, we can compare our high luminosity quasar hosts to the host galaxies of lower-luminosity Type-2 AGNs reported in the literature. Studies of Type-2 AGNs selected using emission line ratios have found that their hosts have star-formation rates below the star-formation main sequence \citep[][]{ellison2016}, and often lie in the green-valley or are quiescent \citep[][]{Kauffmann2003agn, Wild2007, salim2007, Schawinski2007, leslie2015, Lacerda2020}. Comparison with our results would then suggest that stellar populations become younger as AGN luminosity increases \citep[e.g.][]{Kauffmann2003agn, Trump2013, georgantopoulos2023}. This could occur if, for example, more molecular gas leads to more intense star formation and higher AGN luminosity \citep{salim2007, rosario2013b, aird2010, Ni2023}. However, the preference for Type-2 AGNs to inhabit green-valley galaxies may be a selection effect, with strongly star-forming Type-2 AGN host galaxies leading to emission line ratios that preclude them from AGN selection \citep{trump2015}. In such a scenario, the only difference between Type-1 and Type-2 AGN is obscuration and viewing angle. We plan to investigate differences between the stellar population of Type-1 quasars and Type-2 quasar/AGN host galaxies in a forthcoming paper.

\subsection{Post-starburst galaxy - quasar connection}

Figure \ref{fig:quasar_bar} also shows that a large fraction of quasar host galaxies are post-starburst galaxies ($24.9 \pm 0.6\%$) and green-valley galaxies ($17.5 \pm 0.4\%$). The former is especially significant, since it implies an excess of $28\pm1$ times compared to the number of post-starburst galaxies in the control sample. Post-starburst galaxies are known to be rare in the local universe, contributing $\sim$1-2 per cent of the galaxy population \citep[][]{Wild2016, rowlands2018}. Their frequency rises with redshift, but only to 2-3 per cent at $z \sim 0.7$ using similar spectroscopic selection methods \citep{rowlands2018}. Thus, even allowing for the different redshift range of our control sample, post-starburst host galaxies are significantly more common amongst our quasars than inactive galaxies.

An increased preference for post-starburst galaxies, or bursty star formation histories, has been suggested by previous spectroscopic studies of both quasar host galaxies \citep[][]{vdb2006,shi2009, canalizo2013, matsuoka2015, melnick2015, wu2018}, and lower-luminosity AGN host galaxies \citep{Kauffmann2003agn, goto2006, Georgakakis2008, kocevski2009}. Some studies further suggest that the incidence of post-starburst host galaxies increases with AGN luminosity \citep{Kauffmann2003agn, vdb2006, cales2015}. Our results build on this work demonstrating a very strong link between bright quasars and post-starburst host galaxies.  

Several studies have investigated the inverse question of the incidence of AGN activity in post-starburst galaxies, with some studies suggesting a low-to-normal frequency \citep[][]{depropris2014, meusinger2017, Lanz2022, almaini2025}, while others suggest close to ubiquitous presence of (obscured) AGNs in post-starburst galaxies \citep{yan2006, Wild2007,brown2009, yesuf2014, pawlik2018}. The difference almost certainly comes down to selection differences between different post-starburst and AGN samples. Observational studies also suggest there may be a delay in AGN activity following a starburst \citep{davies2007, Schawinski2007, Wild2010}, further solidifying the link between AGN and post-starburst (or green-valley) hosts. While it should be possible to unify these two inverse approaches to identifying a link between post-starburst galaxies and AGN, it is complicated by selection effects on both sides, and will require forward-modelling of AGN and starburst visibility timescales.  

A class of objects known as post-starburst quasars (PSQs) have been studied previously, and are found to have black hole masses and accretion rates indistinguishable from conventional quasars \citep{Cales2013, cales2015}, with hosts showing signatures of past mergers \citep{brotherton1999, canalizo2001, cales2011}. However, these PSQ samples are very different from our post-starburst quasar hosts, as they are typically identified via Balmer (specifically H$\delta$) absorption in their spectra alone \citep{Cales2013, cales2015, melnick2015}. Because A and F stars are present in star-forming galaxies, the selection of PSQs based solely on a cut on H$\delta$ absorption line strength does not exclude galaxies with ongoing star-formation, i.e. galaxies which are not ``post''-burst \citep[e.g.][]{goto2006, wu2018}. Our study provides a census of true post-starburst galaxies among quasars. The addition of a mass-matched control sample of galaxies, analysed using the same spectral decomposition method, ensures that our results are also robust to selection effects of the post-starburst features.

Regardless of the exact nature of the stellar populations of quasar hosts, these pioneering observational studies of PSQs supported the early merger-driven quasar-galaxy co-evolution models \citep{sanders1988}, which were widely replicated by numerical simulations \citep{springel2005, hopkins2008, hopkins2009} and will be discussed in the following subsection. 

\subsection{Implications}

The observed excess of post-starburst quasar host galaxies may imply that such galaxies are particularly conducive to quasar activity. Many studies find that post-starburst galaxies are strongly associated with major gas-rich mergers \citep{ellison2024, sazonova2021, pawlik2018, pawlik2016, wilkinson2022}, and that the post-starburst phase is more common among galaxies with merger signatures than those without \citep{ellison2024, wilkinson2022, li2023}, although simulations suggest that mergers are not a prerequisite for post-starburst galaxies \citep{pawlik2018}. There is also evidence for an increase in the incidence of mergers amongst post-starburst galaxies with stellar mass \citep{pawlik2018,ellison2024}. At stellar masses close to those of our quasar host galaxies, \citet{ellison2024} find that $\sim90$ per cent of post-starburst galaxies show signs of recent gas-rich mergers. The increased frequency of post-starburst galaxies in quasars, along with the high frequency of mergers in massive post-starburst galaxies, points to a connection between gas-rich major mergers and quasar activity.

Observational imaging studies of the connection between AGNs and mergers, via hosts with morphological disturbances, close companions and tidal tails, have yielded mixed results. Some studies find that the most luminous AGNs and quasars are more common among merging galaxies than lower luminosity AGNs \citep{urrutia2008, marian2020, ellison2019}. Other studies have found no connection between quasars and major mergers \citep{villforth2017, hewlett2017}. A large part of the discrepancy between results is due to differences in experimental design. Different techniques, ranging from visual inspection \citep[e.g.][]{kocevski2015}, statistical \citep[e.g.][]{pawlik2016} to machine learning \citep[e.g.][]{bickley2021} are sensitive to different merger features which decay rapidly in brightness after the merger, affecting their detectability \citep{pawlik2016}. Additionally, the quality of imaging data used to identify mergers differs between each study \citep{wilkinson2024}, and identifying faint morphological features, corresponding to mergers, from under a bright quasar point spread function \citep[PSF; e.g.][]{mechtley2016, tang2023} can be quite challenging. Some studies suggest that merger signatures may be associated only with specific populations of quasars, such as red or obscured quasars, for which studies have found correlations between AGN luminosity and mergers \citep[][]{urrutia2008, kocevski2015, glikman2015, dougherty2024}.  These specific sub-populations, rather than the majority of quasars (which we find to be star-forming), may be ideal for studying the major merger-quasar connection.

Our results suggest that supermassive black holes showing quasar activity have two parallel feeding modes. The bulk of quasars have star-forming host galaxies with stellar populations indicative of star-formation histories free of intense time-limited bursts of star-formation. Such objects are likely to show disk-like morphologies \citep[e.g.][]{strateva2001}, and such disk galaxies are likely not to have experienced a major merger in the recent past \citep[][]{martig2012}. In such objects, fuel for SMBH feeding may come from galactic bars \citep{jogee2006}, internal gas instabilities \citep{bornaud2012}, minor mergers \citep{hernquist1995} or gas from supernovae or stellar winds \citep{ciotti2007}. Evidence for this comes from studies finding a lack of mergers among most AGN host galaxies at low-redshifts \citep{cisternas2011, villforth2017}, disk-like morphologies of quasar host galaxies \citep[albeit with significant bulges;][]{cisternas2011, Li2021, tang2023}, and a diversity of feeding mechanisms for SMBHs in simulated galaxies \citep{steinborn2018, choi2024}. 

On the other hand, major mergers might play the leading role in triggering quasar activity in post-starburst host galaxies. \citet{draper2012} and \citet{hirschmann2012} showed that models with parallel SMBH feeding modes provide a good explanation for the form of AGN luminosity functions. While quasars in star-forming host galaxies may represent the bulk of SMBH growth, post-starburst quasars may represent short bursts of significant SMBH growth - an extension of the results of \citet{Wild2007}. A prediction of this quasar feeding dichotomy is a higher AGN luminosity of merger-driven, post-starburst quasars compared to non-merger star-forming quasars \citep{hopkins2009, storchi2019}. We plan to test the proposed parallel modes of quasar feeding in future work.

\begin{table}
\centering
\caption{Projected merger fractions of quasar host galaxies in our sample, based on their recent star formation histories. From left to right: (1) $f_{m}$ is the fraction of quasar hosts that might have experienced a recent major merger. (2) $f_{s}$ $(=1-f_{m})$ is the fraction of quasars that are not expected to have experienced a recent major merger (i.e. secularly triggered). Errors are from adding standard deviations on host type fractions in quadrature.}
\label{tab:merge_frac}
\begin{tabular}{lcc}
\hline
 & Merger fraction & Secular fraction\\
 & $(f_{m})$ & $(f_{s})$\\
\hline
Assuming all GVs are secularly fed & $0.31\pm0.01$ & $0.69\pm0.01$\\
Assuming an equal split & $0.40\pm0.01$ & $0.60\pm0.01$\\
Assuming all GVs are merger fed & $0.49\pm0.01$ & $0.51\pm0.01$\\
\hline
\end{tabular}
\end{table}

Given the difficulty in determining the morphologies of high-luminosity AGN host galaxies \citep[e.g.][]{mechtley2016, zhao2022, tang2023}, we can use simplifying assumptions to compile our results quantitatively into merger fractions for quasars. To calculate the fraction of quasars fed by mergers ($f_{m}$) and fed by secular processes ($f_{s}$), we first assume that all our (high-stellar mass) post-starburst and starburst quasar hosts have experienced a recent ($<1$ Gyr ago) major merger \citep[][]{ellison2024}. By contrast, we assume that none of the star-forming (i.e. non-burst/smooth/continuous SFHs) quasar host galaxies have experienced a recent major merger. Finally, we assume that quasars in quiescent host galaxies are fed by secular processes like stellar winds, owing to low gas fractions \citep[][]{Ni2023}. We adopt three extreme cases for green-valley quasars - a) all are fed by major mergers, b) all are fed by secular processes, c) an equal split between the two. This gives us a range of $f_{m}$ and $f_{s}$ values, which are shown in the Table \ref{tab:merge_frac}. We find that $f_{m}$ ranges from $\sim0.3$ to $0.5$, conversely $f_{s}$ ranges from $0.5$ to $0.7$. These merger fractions are consistent with those presented by \citet{marian2020} ($f_{m} = 0.41 \pm 0.12$) and \citet{ellison2019} ($f_{m} = 0.37$) for high luminosity optically-selected AGNs. These results quantify our previous assertion, that while mergers are conducive to quasar activity, quasars are not primarily triggered by mergers \citep[][]{ellison2019, marian2020}. 

The large number of star-forming quasar host galaxies implies that a single episode of quasar activity does not always correspond to a quenched host galaxy. While such a result may not be at odds with AGN feedback \citep[see][]{Ward2022}, this contrasts studies that support AGN feedback models owing to their hosts being in the green-valley \citep{Schawinski2007}. While older studies argued that the complete exhaustion of gas by star formation alone produces post-starburst galaxies \citep[e.g.][]{Wild2010, Hopkins2012}, with the quasar activity ensuring sustained quenching, this may have been due to the comparatively low resolution of the simulations at that time \citep{zheng2020}. Most modern simulations require some form of AGN feedback to produce rapidly quenched post-starburst galaxies \citep[][]{rm2019, zheng2020, lotz2021}. The existence of cold gas in and around many post-starburst galaxies \citep[][]{zwaan2013, french2015, rowlands2015, ellison2025}, however, challenges the need for complete exhaustion of gas supplies. It would be interesting to compare the properties of quenching and star forming quasar hosts to see whether a direct link between mergers and quasar activity can be found. In this regards, mergers may actually play a key role in determining which galaxies experience AGN feedback \citep{davies2022}.

\section{Conclusion}

We introduced Mean-Field Independent Component Analysis (MFICA) to decompose quasar spectra. We applied MFICA to samples of SDSS DR7 galaxy and quasar spectra, obtaining data-driven templates (components) that reconstruct these spectra. The galaxy components resemble stellar populations of galaxies, and their fractional contributions allow estimation of the stellar populations, and therefore recent star-formation histories (SFHs) of galaxies. Using mock spectra, we show that MFICA can recover the fraction of galaxy light embedded in quasar spectra ($f_{gal}$) and, under the limits of $f_{gal} \geq 0.2$ and $S/N \geq 10$, can determine the stellar populations of quasar host galaxies.

We applied MFICA decomposition to a sample of SDSS DR7 quasars in the redshift range $0.16 \leq z \leq 0.76$, with the following main science results: 

\begin{enumerate}
    \item Quasars are hosted by massive galaxies with a median fibre stellar mass of $10^{10.8}$ $\rm M_{\sun}$ and total stellar mass of $\sim10^{11.3}$ $\rm M_{\sun}$. This agrees with e.g. clustering results that find luminous AGN to be hosted by massive galaxies. 
    \item Around 53 per cent of quasar host galaxies are star-forming or starburst galaxies, compared to 14 per cent of a mass-matched control sample of galaxies. Conversely, $\sim5$ per cent of quasar host galaxies are quiscent/red, compared to 74 per cent of control galaxies. This indicates that luminous AGN are not simply a random sub-sample of the galaxy population, and young-stellar populations and AGN activity are directly or indirectly causally linked. 	
    \item Around 25 per cent of all quasars are hosted by post-starburst galaxies, an excess of $28\pm1$ times compared to the number of post-starburst galaxies in the control sample. This may also be viewed as a $7.3\pm0.4$ excess of post-starbursts among quasars, for the same number of star-forming galaxies. Based on the known connection between massive post-starburst galaxies and major mergers, we argue that the connection between post-starburst hosts and luminous quasars may be mediated by major mergers.
\end{enumerate}

Our results fit into a scenario whereby quasars have two parallel feeding modes: i) a post-starburst mode of rapid black hole growth driven by mergers, and ii) a star-forming mode driven by secular mechanisms like minor mergers and bars, where black hole growth is in step with star-formation owing to a common gas supply. Under simplifying assumptions, this picture leads to a merger fraction of 30-50 per cent for quasars that is consistent with morphological studies in literature.

Finally, while our results show that a single episode of quasar activity does not typically coincide with a quenching of star formation in the galaxy, the rapid truncation of star formation in quasars with post-starburst hosts warrants further investigation given the significant implications for understanding galaxy quenching more generally.

\section*{Acknowledgements}

We would like to thank the anonymous referee for their careful reading and comments which have improved the work presented here. SDK thanks Ho-Hin Leung, Alfie Russell and Rita Tojeiro for their help and constructive feedback during the early parts of this project. VW acknowledges the Science and Technologies Facilities Council (ST/Y00275X/1) and Leverhulme Research Fellowship (RF-2024-589/4). 

Funding for the SDSS and SDSS-II has been provided by the Alfred P. Sloan Foundation, the Participating Institutions, the National Science Foundation, the U.S. Department of Energy, the National Aeronautics and Space Administration, the Japanese Monbukagakusho, the Max Planck Society, and the Higher Education Funding Council for England. The SDSS Web Site is http://www.sdss.org/.

The SDSS is managed by the Astrophysical Research Consortium for the Participating Institutions. The Participating Institutions are the American Museum of Natural History, Astrophysical Institute Potsdam, University of Basel, University of Cambridge, Case Western Reserve University, University of Chicago, Drexel University, Fermilab, the Institute for Advanced Study, the Japan Participation Group, Johns Hopkins University, the Joint Institute for Nuclear Astrophysics, the Kavli Institute for Particle Astrophysics and Cosmology, the Korean Scientist Group, the Chinese Academy of Sciences (LAMOST), Los Alamos National Laboratory, the Max-Planck-Institute for Astronomy (MPIA), the Max-Planck-Institute for Astrophysics (MPA), New Mexico State University, Ohio State University, University of Pittsburgh, University of Portsmouth, Princeton University, the United States Naval Observatory, and the University of Washington.

\section*{Data Availability}

Data used in this paper is publicly available from the SDSS DR7 website \url{http://classic.sdss.org/dr7/}. OH skyline-subtracted SDSS DR7 spectra are available from \url{http://www.sdss.jhu.edu/skypca/spSpec}. The MFICA templates are available online via the supplementary material held by the publisher.



\bibliographystyle{mnras}
\bibliography{mnras_template} 




\appendix

\section{Effect of varying S/N and $f_{gal}$ on component fractions}\label{sn_fgal_eff}

\begin{figure*}
\centering
\includegraphics[width=\textwidth]{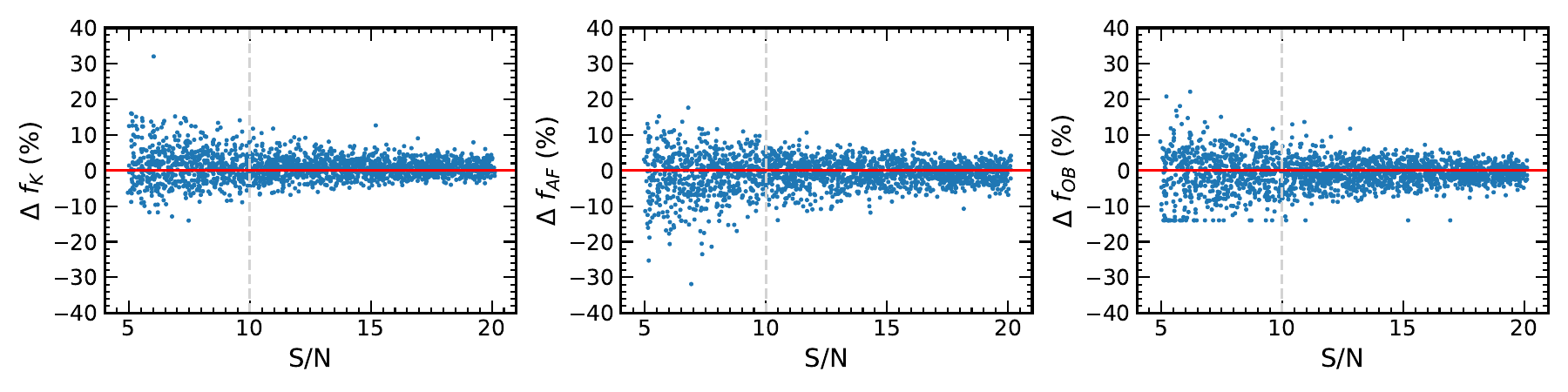}
\caption{Scatter plots showing the per cent difference in the component fractions from decomposing mock spectra with a fixed $f_{gal}$ = 0.2 and star-forming host galaxy compared to the true component fractions, as a function of the median S/N ratio of the spectra. The percentages are measured with respect to the dynamical ranges of the individual component fractions. From left to right we show the K, AF and OB-component fractions. The dashed grey line represents the S/N$= 10$ lower limit used in our study.}
\label{fig:weight_vs_snr}
\end{figure*}

\begin{figure*}
\centering
\includegraphics[width=\textwidth]{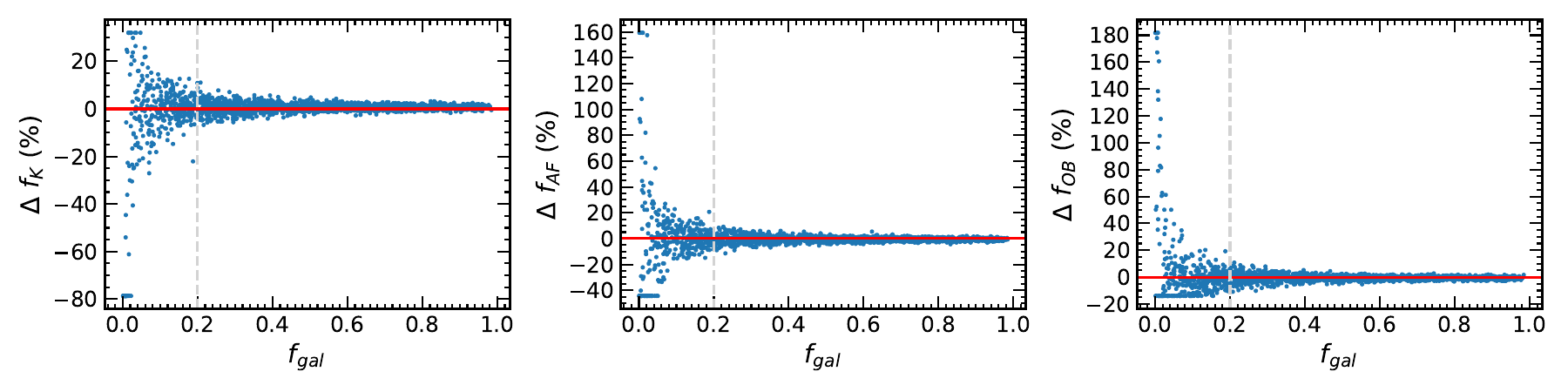}
\caption{Scatter plots showing the per cent difference in the component fractions from decomposing mock spectra with a fixed median S/N$=10$ and star-forming host galaxy compared to the true component fractions, as a function of $f_{gal}$. The percentages are measured with respect to the dynamical ranges of the individual component fractions. From left to right we show the K, AF and OB-component fractions. The dashed grey line represents the $f_{gal} = 0.2$ lower limit used in our study.}
\label{fig:weight_vs_fgal}
\end{figure*}

In this section, we describe how the recovery of galaxy component fraction from mock spectra varies as a function of $f_{gal}$ and the median S/N ratio of the spectra.

To quantify the dependence of component fraction recovery on the median S/N of spectra, we create 2000 mock spectra with a star-forming host galaxy and randomly chosen quasar spectra (see Section \ref{mock_decomp} for more details). We assign each spectrum a random median S/N between 5 and 20, while fixing $f_{gal}=0.2$. In Fig.~\ref{fig:weight_vs_snr}, we plot the per cent difference in the component fractions from decomposing the mock spectra compared to their actual values against the median S/N of the spectra. The percentage differences have been calculated with respect to the dynamic ranges of the component fractions. It is clear that as the median S/N increases, the recovery of the host galaxy component fractions improves, albeit weakly.

To quantify the dependence of component fraction recovery on $f_{gal}$, we create 2000 mock spectra with a star-forming host galaxy and randomly chosen quasar spectra. We assign each spectrum a random $f_{gal}$ between 0 and 1, and fix S/N $=10$. In Fig.~\ref{fig:weight_vs_fgal}, we plot the per cent difference in component fractions with respect to the dynamic ranges of the component fractions, against $f_{gal}$. The recovery of the weights improves dramatically with an increase in $f_{gal}$. This improvement is more substantial than a change in median S/N, highlighting that the choice of $f_{gal}$ is more important in determining the performance of MFICA decomposition.


\bsp	
\label{lastpage}
\end{document}